\documentclass[12pt, twocolumn,showpacs,preprintnumbers,aps,prl,reprint,groupedaddress]{revtex4-1}

\usepackage{amssymb}
\usepackage{amsmath}
\usepackage{subfigure}
\usepackage{graphicx}
\usepackage{color}

\newcommand{\subfigimg}[3][,]{%
  \setbox1=\hbox{\includegraphics[#1]{#3}}
  \leavevmode\rlap{\usebox1}
  \rlap{\hspace*{3pt}\raisebox{\dimexpr\ht1-1\baselineskip}{#2}}
  \phantom{\usebox1}
}

\newcommand{\subfigimgloc}[5][,]{%
  \setbox1=\hbox{\includegraphics[#1]{#5}}
  \leavevmode\rlap{\usebox1}
  \rlap{\hspace*{#3pt}\raisebox{\dimexpr\ht1-#4\baselineskip}{#2}}
  \phantom{\usebox1}
}

\newcommand{\subfigimgloctwice}[8][]{%
  \setbox1=\hbox{\includegraphics[#1]{#8}}
  \leavevmode\rlap{\usebox1}
  \rlap{\hspace*{#3pt}\raisebox{\dimexpr\ht1-#4\baselineskip}{#2}}
  \rlap{\hspace*{#6pt}\raisebox{\dimexpr\ht1-#7\baselineskip}{#5}}
  \phantom{\usebox1}
}

\newcommand{\wb}[1]{\colorbox{white}{#1}}

\begin{document}
\preprint{}

\title{Instabilities and Topological Defects in Active Nematics}

\author{Sumesh P. Thampi}
\affiliation{The Rudolf Peierls Centre for Theoretical Physics, 1 Keble Road, Oxford, OX1 3NP, UK}

\author{Ramin Golestanian}
\affiliation{The Rudolf Peierls Centre for Theoretical Physics, 1 Keble Road, Oxford, OX1 3NP, UK}

\author{Julia M. Yeomans}
\affiliation{The Rudolf Peierls Centre for Theoretical Physics, 1 Keble Road, Oxford, OX1 3NP, UK}
\email[]{j.yeomans1@physics.ox.ac.uk}
\homepage[]{http://www-thphys.physics.ox.ac.uk/people/JuliaYeomans/}


\begin{abstract}{We study a continuum model of an extensile active nematic to show that mesoscale turbulence develops in two stages: (i) ordered regions undergo an intrinsic hydrodynamic instability generating walls, lines of stong bend deformations, (ii) the walls relax by forming oppositely charged pairs of defects. Both creation and annihilation of defect pairs reinstate nematic regions which undergo further instabilities, leading to a dynamic steady state. We compare this with the development of active turbulence in a contractile active nematic.}
\end{abstract}

\maketitle

\begin{figure}
\subfigure{\subfigimgloc[trim = 115 45 90 30, clip, width=0.49\linewidth]{\colorbox{white}{(a)}}{3}{1}{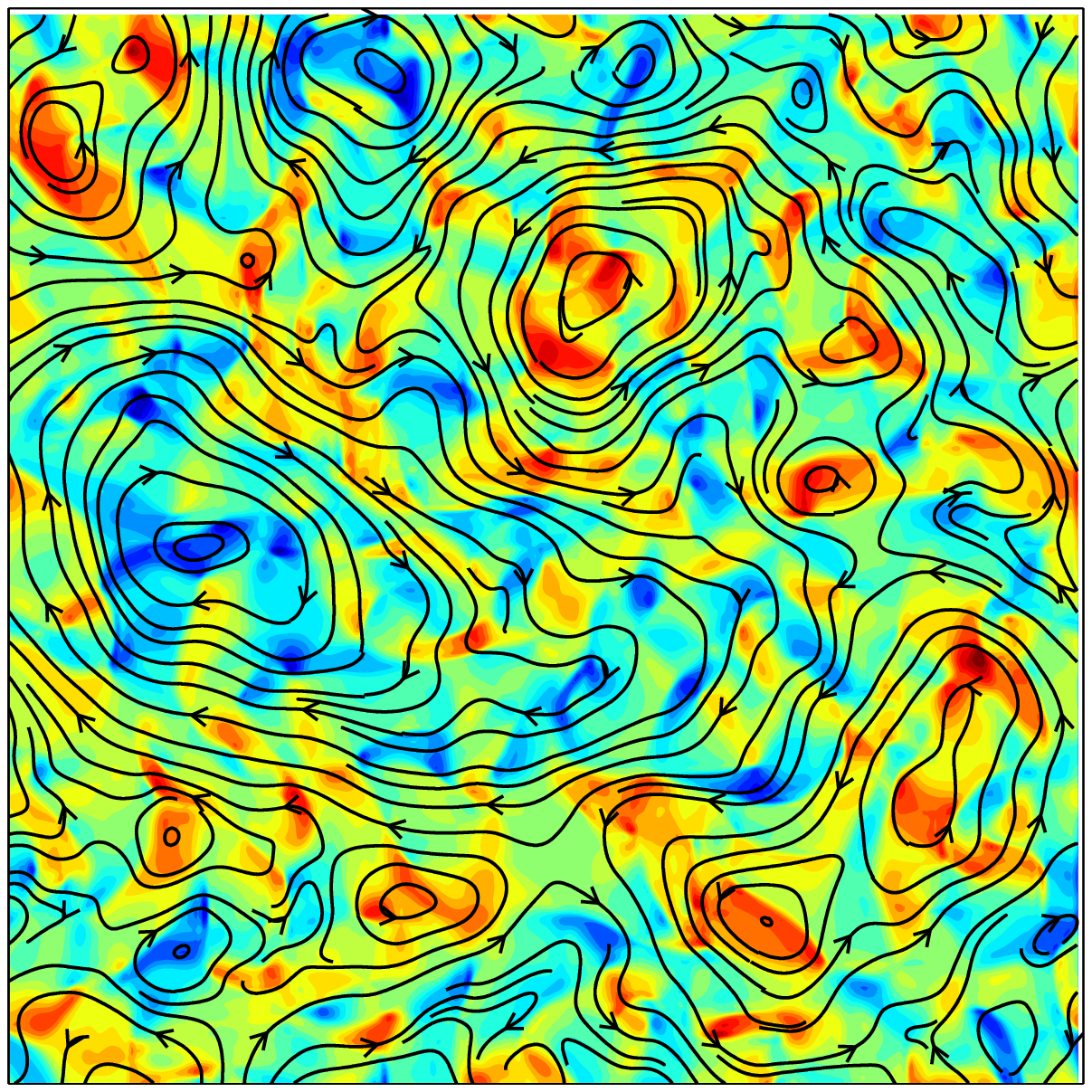}\label{fig:fullfledgev}}
\subfigure{\subfigimgloc[trim = 115 45 90 30, clip, width=0.49\linewidth]{\colorbox{white}{(b)}}{3}{1}{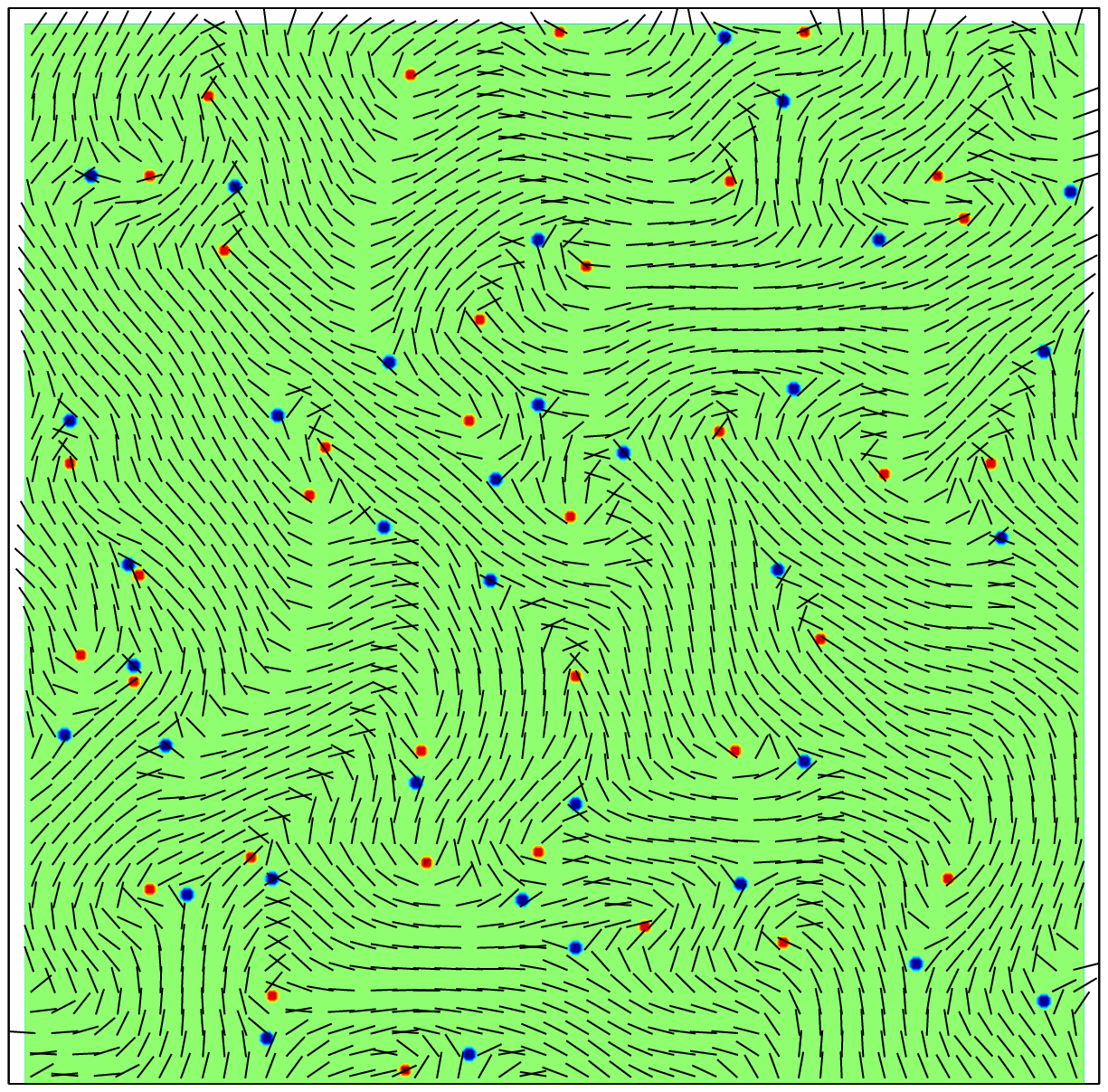}\label{fig:fullfledged}}
\caption{Simulations of fully developed active turbulence. (a) streamlines and vorticity field, with red and blue colouring corresponding to high $+$ve and $-$ve vorticity respectively, (b) director field and topological defects ($+{1}/{2}$, red; $-{1}/{2}$, blue). Panel~(b) shows the left bottom quarter of the domain of (a). }
\label{fig:fullfledge}
\end{figure}

{\em Introduction:} Dense active systems  have generated much interest due to  the novel properties that arise because they operate out of thermodynamic equilibrium \cite{Sriram2010, Ganesh2011, Marchetti2013}. There are many different active suspensions, comprising individual components that can differ widely in their characteristic length scales. Examples include mixtures of molecular motors and microtubules or actin, cells and bacteria, vibrating granular rods, and schools of fish \cite{Sriram2010, Ganesh2011, Marchetti2013, Bausch2010, Chate2012, Dogic2012, Julia2012, Narayan2007}. At high densities  active systems often exist in a state where the velocity field is highly disordered, with a continually changing pattern of vortices, see fig.~\ref{fig:fullfledge}. The turbulent appearance of the flow is, at first sight, surprising because the active suspensions usually correspond to a low Reynolds number regime. However detailed properties, such as scaling laws, are very different to inertial turbulence \cite{Julia2012, Jorn2013}. 

For active materials with hydrodynamic interactions a linear stability analysis shows that the nematic state is unstable to fluctuations \cite{Sriram2002, Madan2007, Scott2009, Joanny2005}. However, the path to well-developed mesoscale turbulence is not yet clear \cite{Mahadevan2011}. Moreover the extent to which the evolution of the turbulent state is generic between different active systems is still to be understood. Here we contribute to answering these questions by demonstrating the route through which the turbulent state is reached in a 2D active nematic.

In addition to hydrodynamic instabilities there is now considerable evidence that topological defects play a role in determining the dynamics of active systems. The  presence of topological defects in an active system with  nematic symmetry has been demonstrated in experiments using 2D suspensions of microtubule bundles and kinesin molecular motors \cite{Dogic2012}. Simulations show that  such defects are strongly associated with vorticity generation in extensile active nematics \cite{ourprl2013}.  Moreover, defects have been identified in dry active matter experiments\cite{Kremkemer2000, Narayan2007} and in active systems with polar symmetry \cite{Kruse2004, Bausch2013}. 

Generally, in passive systems, oppositely charged defects formed, say by a quench, attract and annihilate each other and thus defects continuously disappear from a system as it approaches equilibrium. By contrast, because of the continuous input of energy, defects in active systems can be formed in pairs and subsequently move apart \cite{Giomi2013, ourprl2013} giving rise to a steady state where topological defects are continually being created and destroyed.

In this letter, we perform simulations of active nematics to study the onset of active turbulence in detail. Two physical mechanisms are shown to be relevant to setting up and maintaining the flow field. The first is the formation of lines of kinks in the director field, which we shall term \textit{walls}, and which arise directly from the hydrodynamic instability of the active nematic. The second is the `unzipping' of the walls by the formation of topological defects, a process that we show can be driven either by flow or by relaxation of the excess elastic free energy in the wall. We shall first introduce the model and then describe the onset of active turbulence. Wall formation and defect formation will each be discussed in more detail. We concentrate primarily on extensile systems, but discuss contractile nematics  later in the paper.

\begin{figure*}
\centering
\subfigure{\subfigimg[trim = 115 45 100 30, clip, width=0.19\linewidth]{\colorbox{white}{(a)}}{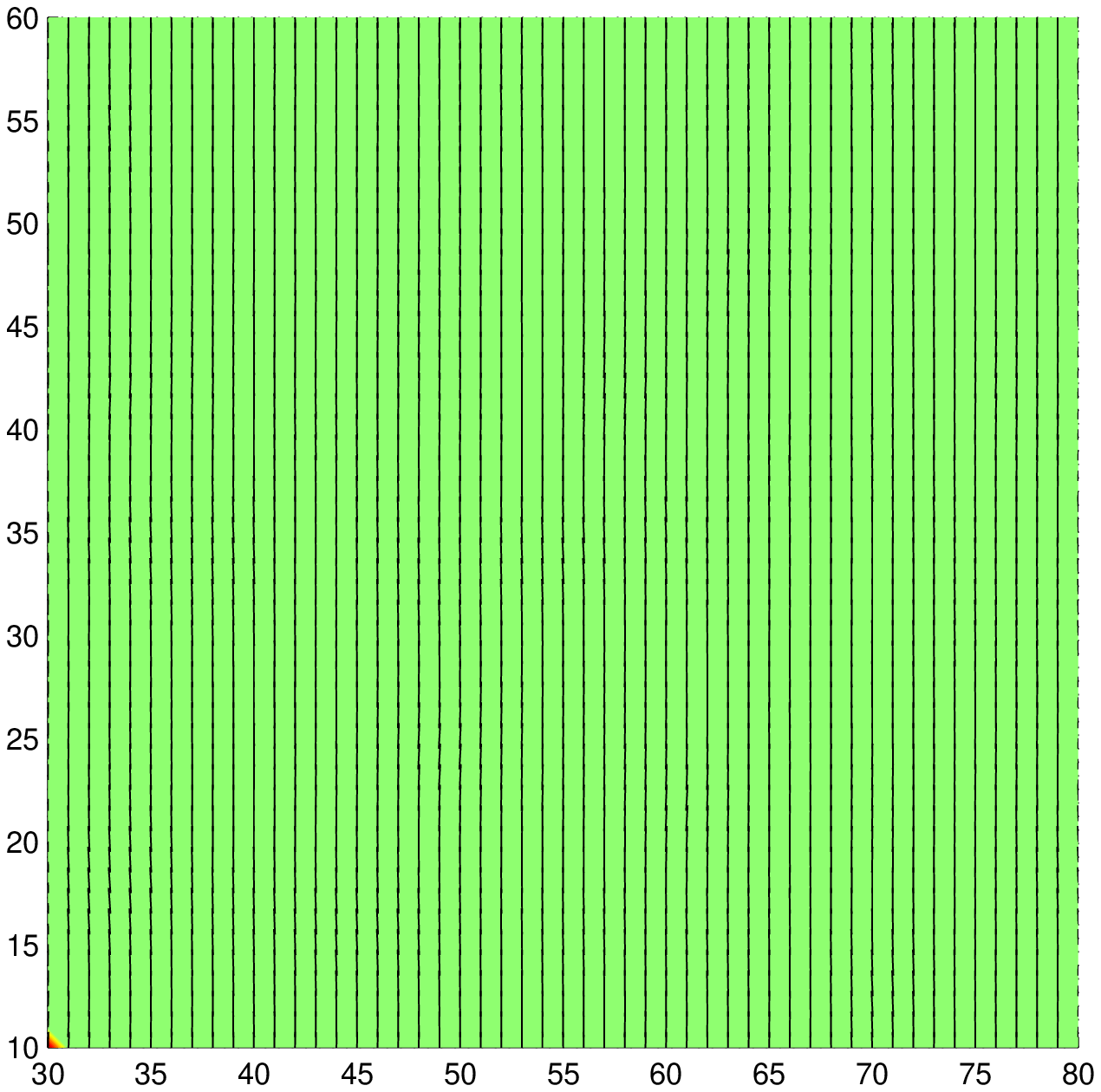}\label{fig:startupt1}}
\subfigure{\subfigimg[trim = 115 45 100 30, clip, width=0.19\linewidth]{\colorbox{white}{(b)}}{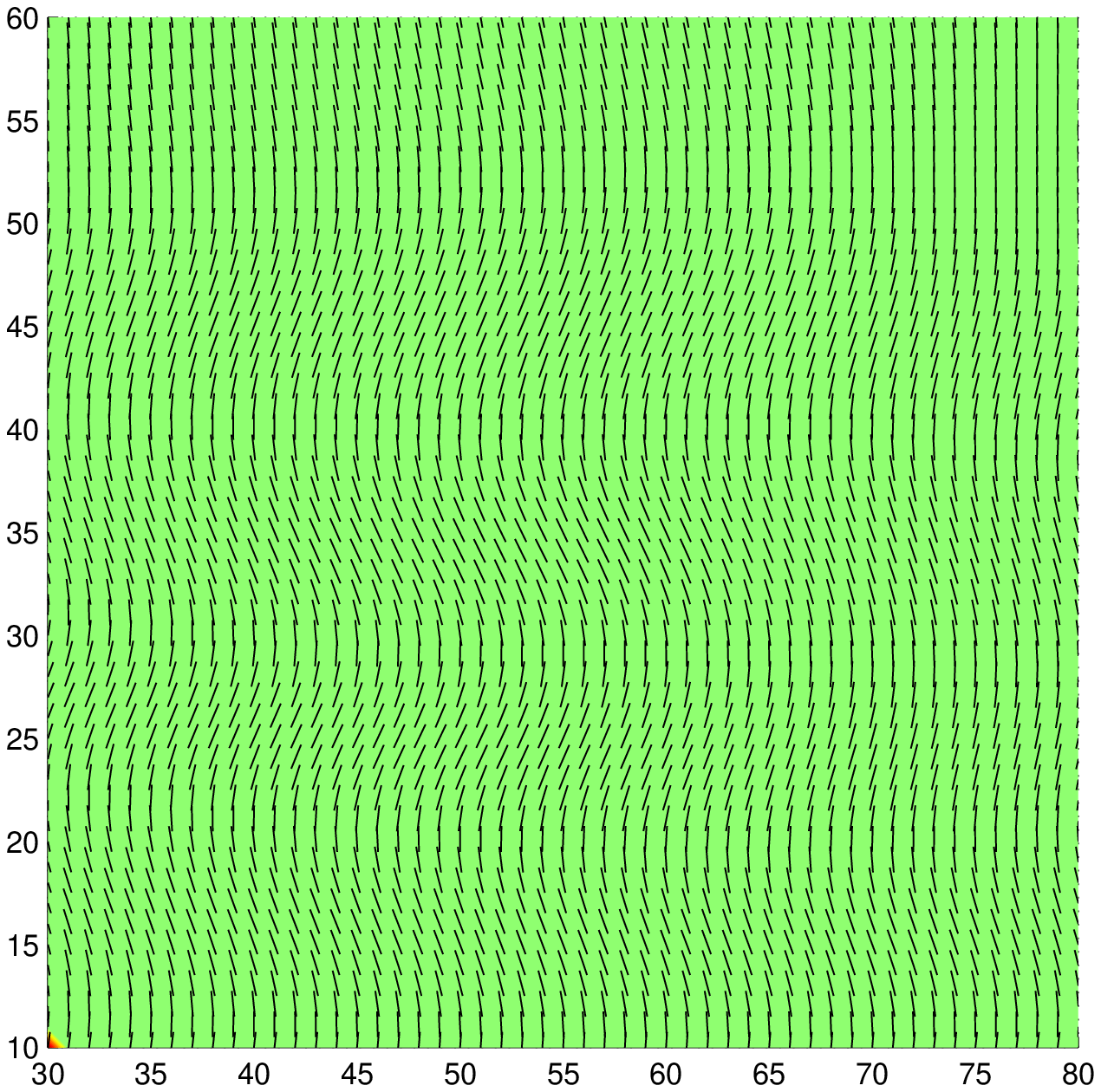}\label{fig:startupt2}}
\subfigure{\subfigimg[trim = 115 45 100 30, clip, width=0.19\linewidth]{\colorbox{white}{(c)}}{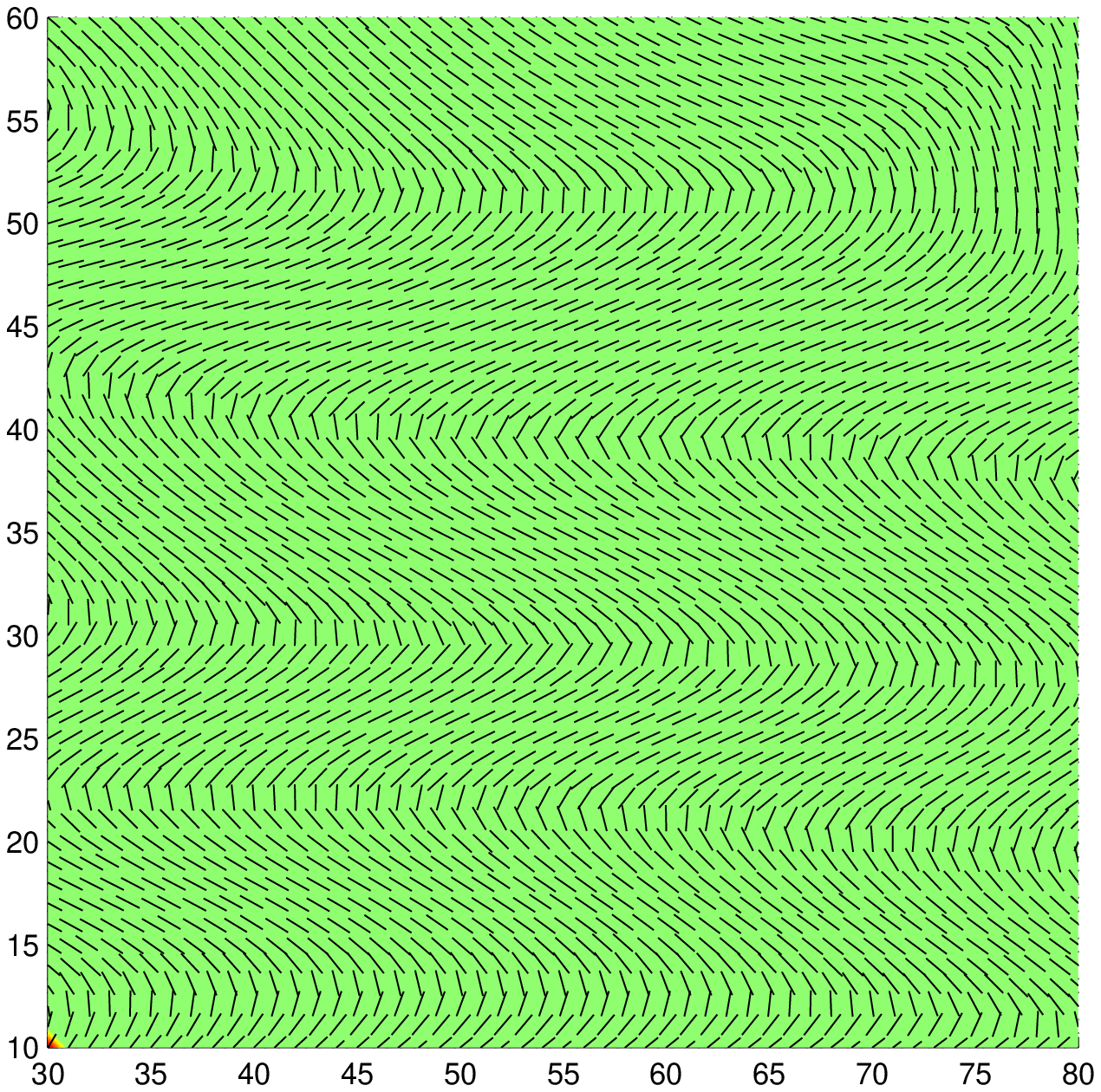}\label{fig:startupt3}}
\subfigure{\subfigimg[trim = 115 45 100 30, clip, width=0.19\linewidth]{\colorbox{white}{(d)}}{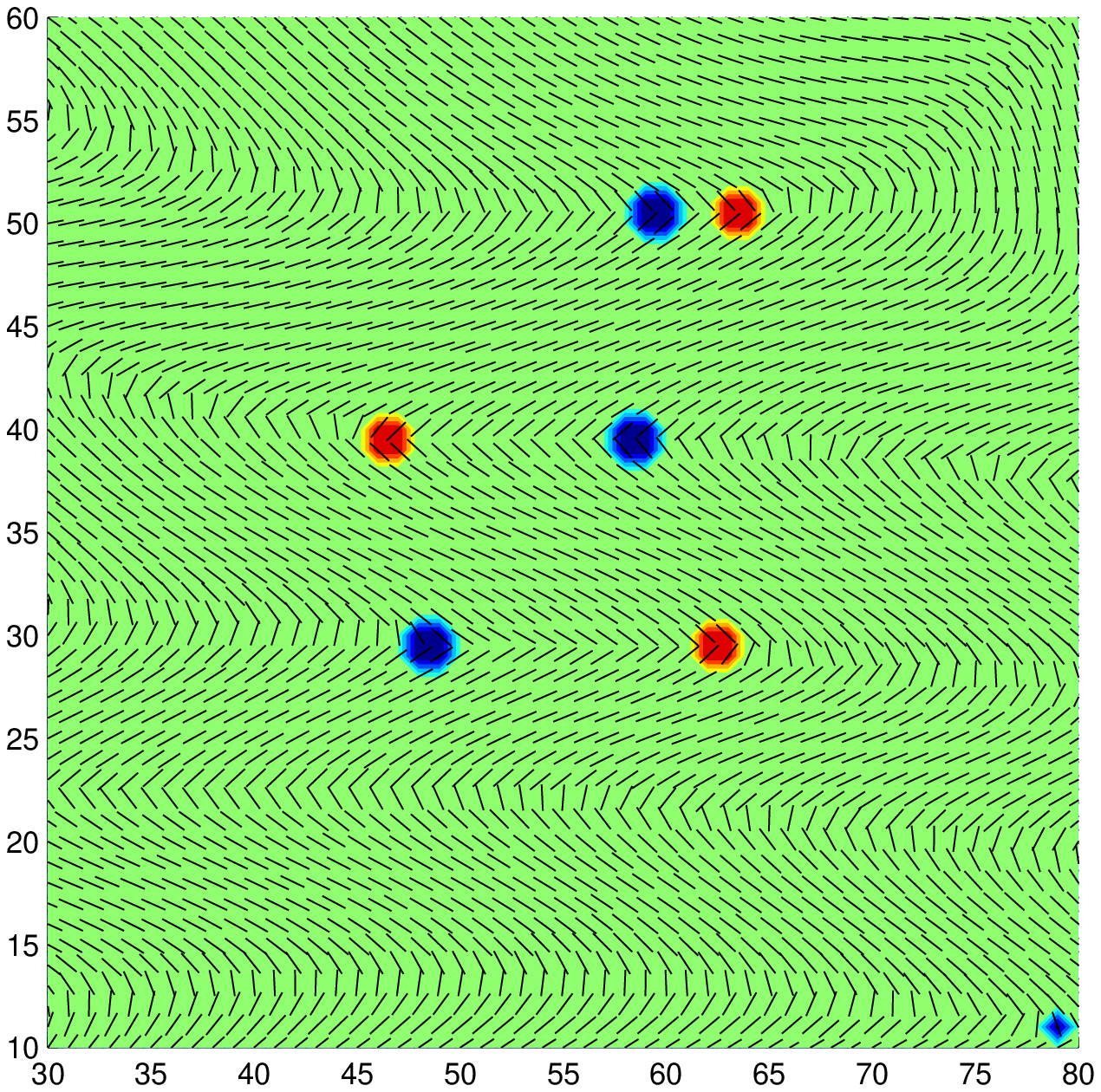}\label{fig:startupt4}}
\subfigure{\subfigimg[trim = 115 45 100 30, clip, width=0.19\linewidth]{\colorbox{white}{(e)}}{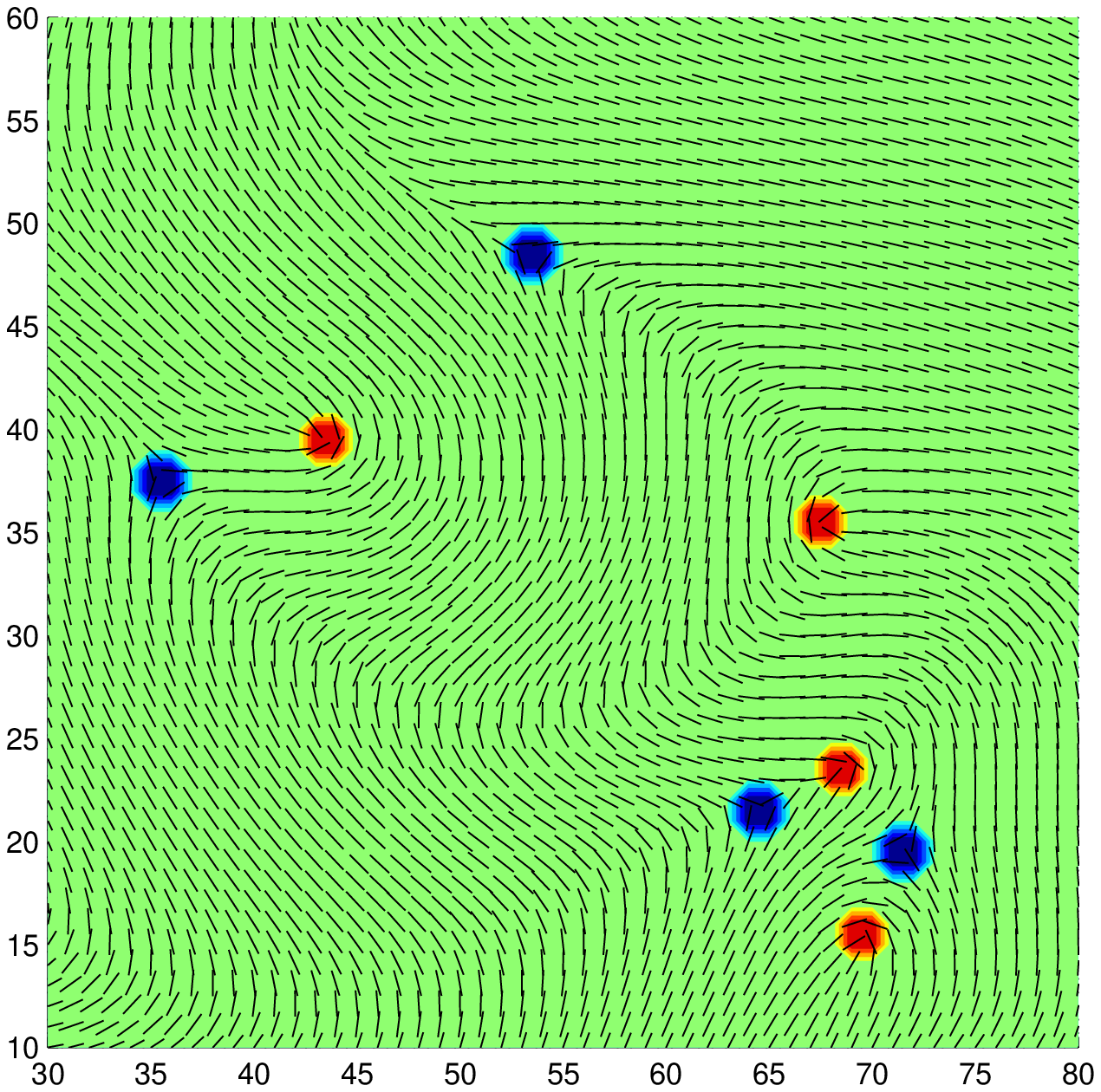}\label{fig:startupt5}}
\caption{Snapshots at successive times of the director field (dashed lines) and $+1/2$ and $-1/2$ defects (red and blue respectively) during the development of active turbulence from an ordered nematic state for an extensile system. Two stages are seen: (b),(c) walls are formed, (d),(e) local nematic order is restored by the formation of pairs of oppositely charged defects.}
\label{fig:startup}
\end{figure*}

{\em Equations of motion:} We consider an active nematic suspension. The evolution equation for the order parameter tensor $\mathbf{Q}$ is a standard equation in liquid crystal hydrodynamics \cite{Berisbook, DeGennesBook}:
\begin{equation}
(\partial_t + u_k \partial_k) Q_{ij} - S_{ij} = \Gamma H_{ij},
\label{eqn:lc}
\end{equation}
where $\mathbf{u}$ is the velocity field and, because the nematic order can respond to  shear flow, the advection term is generalised to
\begin{align}
S_{ij} =& (\lambda E_{ik} + \Omega_{ik})(Q_{kj} + \delta_{kj}/3) + (Q_{ik} + \delta_{ik}/3)\nonumber\\
& (\lambda E_{kj} - \Omega_{kj}) - 2 \lambda (Q_{ij} + \delta_{ij}/3)(Q_{kl}\partial_k u_l),\nonumber
\end{align}
where the strain rate tensor, $E_{ij} = (\partial_i u_j + \partial_j u_i)/2$ and the vorticity tensor, $\Omega_{ij} = (\partial_j u_i - \partial_i u_j)/2$. 
The alignment parameter is chosen as $\lambda=0.7$. 
The relaxation of $\mathbf{Q}$ is related to the molecular field $H_{ij} = -\delta \mathcal{F}/ \delta Q_{ij} + (\delta_{ij}/3) {\rm Tr} (\delta \mathcal{F}/ \delta Q_{kl})$ by the constant of proportionality $\Gamma$, the rotational diffusivity. The standard Landau-de Gennes free energy functional, within the single elastic constant, $K$, approximation,
\begin{align}
\mathcal{F} &= \frac{K}{2} (\partial_k Q_{ij})^2 \nonumber\\
&+ \frac{A}{2} Q_{ij} Q_{ji} + \frac{B}{3} Q_{ij} Q_{jk} Q_{ki} + \frac{C}{4} (Q_{ij} Q_{ji})^2 \nonumber
\end{align}
is used to determine the molecular field, $\mathbf{H}$.
Here, $A, B$ and $C$ are material constants. 

The velocity field obeys the equations of motion
\begin{equation}
\begin{aligned}
\nabla \cdot \mathbf{u} &= 0;~
\rho (\partial_t + u_k \partial_k) u_i &= \partial_j \Pi_{ij}.
\label{eqn:ns}
\end{aligned}
\end{equation}
where $\rho$ is the fluid density. The stress tensor $\mathbf{\Pi}$ incorporates two terms which appear in the hydrodynamic equations describing passive liquid crystals
\begin{itemize}
\item the viscous stress, $\Pi_{ij}^{viscous} = 2 \mu E_{ij}$,
\item the passive stress,
$\Pi_{ij}^{passive}=-P\delta_{ij} + 2 \lambda(Q_{ij} + \delta_{ij}/3) (Q_{kl} H_{lk})
-\lambda H_{ik} (Q_{kj} + \delta_{kj}/3)  - \lambda (Q_{ik} + \delta_{ik}/3) H_{kj}
-\partial_i Q_{kl} \frac{\delta \mathcal{F}}{\delta \partial_j Q_{lk}} + Q_{ik}H_{kj} - H_{ik} Q_{kj}$
\end{itemize}
where  $P$ is the pressure and $\mu$ is the viscosity. The  activity is imparted by incorporating
\begin{itemize}
\item the active stress, $\Pi_{ij}^{active} = -\zeta Q_{ij}$ introduced in \cite{Sriram2002},
\end{itemize}
where  $\zeta$ is the coefficient controlling the strength of the activity. This term implies that any gradient in $\mathbf{Q}$ will produce a flow field, which is extensile for $\zeta>0$ and contractile for $\zeta<0$. 
More details of the model can be found in \cite{Berisbook, DeGennesBook, Davide2007, Henrich2010}.

The governing equations (\ref{eqn:lc}) and (\ref{eqn:ns}) form a coupled system which we solve using a 
hybrid lattice Boltzmann algorithm \cite{Davide2007, Suzanne2011}. The parameters used are $\Gamma=0.34$, $A=0.0$, $B=-0.3$, $C=0.3$, $K=0.02$, $\mu=2/3$ and $\zeta=0.0125$ unless specified otherwise. These parameters are non-dimensionalised in lattice units where discrete space and time steps are chosen as unity. Depending on the material of interest (cytoskeletal filaments or bacterial suspensions) appropriate scales can be chosen to convert them to physical units \cite{Cates2008, Henrich2010, ourprl2013}.


{\em The onset of active turbulence:}
 Figure~\ref{fig:startup}  illustrates how an active suspension with extensile stress undergoes the transition from a nematic to  a turbulent state which generates and sustains vortical structures in the flow field.

The nematic initial condition is shown in fig.~\ref{fig:startupt1}. Any small bend fluctuation in an ordered active nematic is reinforced by local shear leading to a hydrodynamic instability \cite{Sriram2002, Madan2007, Scott2009}. The linear stability analysis predicts that long wavelength modes are unstable, and the waves of bend deformations shown in fig.~\ref{fig:startupt2} are a consequence of the dominance of the most unstable mode. The bends then sharpen (fig.~\ref{fig:startupt3}) \cite{Madan2007, Scott2009} because the shear flow associated with gradients in the director field acts to further tilt the director to form approximately equispaced lines of kinks similar to the observations in \cite{Mahadevan2011}. We shall refer to these as walls.  Similar structures are observed in passive liquid crystals where they form due to the imposed boundary conditions or the application external forces \cite{Rey2002,Lozar2005}.
 By contrast,  in active matter the wall formation is   internally driven by the flow field generated by the active stress. 

These structures are similar to those formed by an active nematic confined to a bounded channel where, as the activity is increased,  there is a spontaneous symmetry breaking to a state where a kinked director field produces a net flow  \cite{Joanny2005, Davide2007}. However, there the solid boundaries impose 
 no-slip boundary conditions or fix the anchoring of the director field. This is not the case for our fully 2D system and consequently no steady state flow or director field is obtained. Instead, the director continues to tilt until pairs of oppositely charged defects are created as shown in fig.~\ref{fig:startupt4}. This spontaneous formation of defects occurs at the points where the bend is strongest due to local noise or interactions with neighbouring walls. The defects are strong sources of vorticity and move in their own or in the ambient flow \cite{Julia2002, Giomi2013, Pismen2013}. This further destroys the striated structure in the director field which soon looks disordered as shown in fig.~\ref{fig:startupt5} and on a larger scale in fig.~\ref{fig:fullfledge}. The system reaches a dynamic steady state where walls are continually formed and then decay through defect formation. The defect number also saturates because they are advected by the flow field and annihilate if they encounter a defect of the opposite charge.

Thus two distinct processes contribute to the dynamics of active turbulence in extensile nematics; wall formation, and defect formation and annihilation which acts to remove the walls. We next analyse each of these in more detail.

\begin{figure}
\begin{minipage}{0.39\linewidth}
\centering
\subfigure[\ $\zeta=0.005$]{\includegraphics[trim = 70 140 100 30, clip, width=\linewidth]{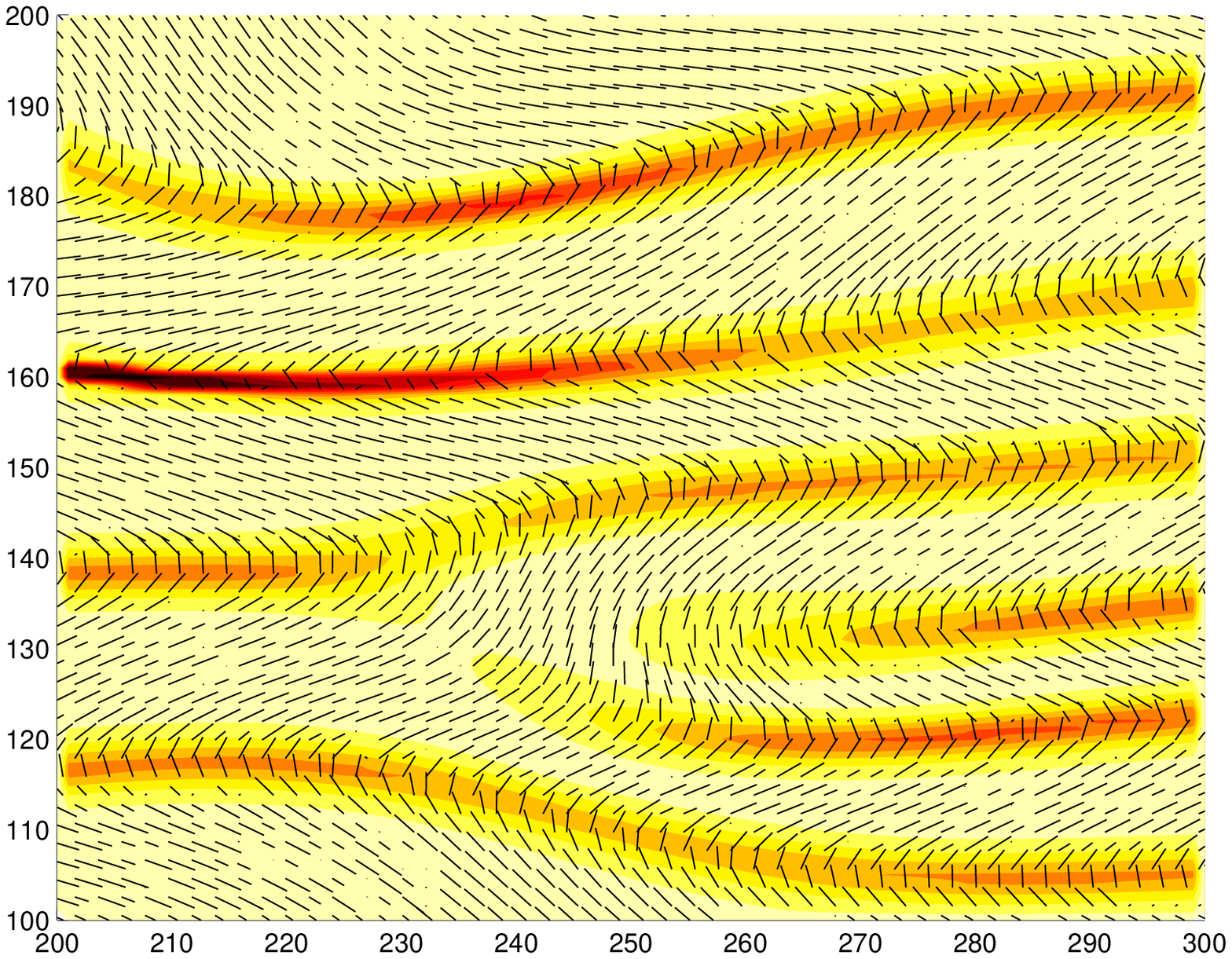}\label{fig:startupzetaa}}\\
\subfigure[\ $\zeta=0.0125$]{\includegraphics[trim = 70 140 100 30, clip, width=\linewidth]{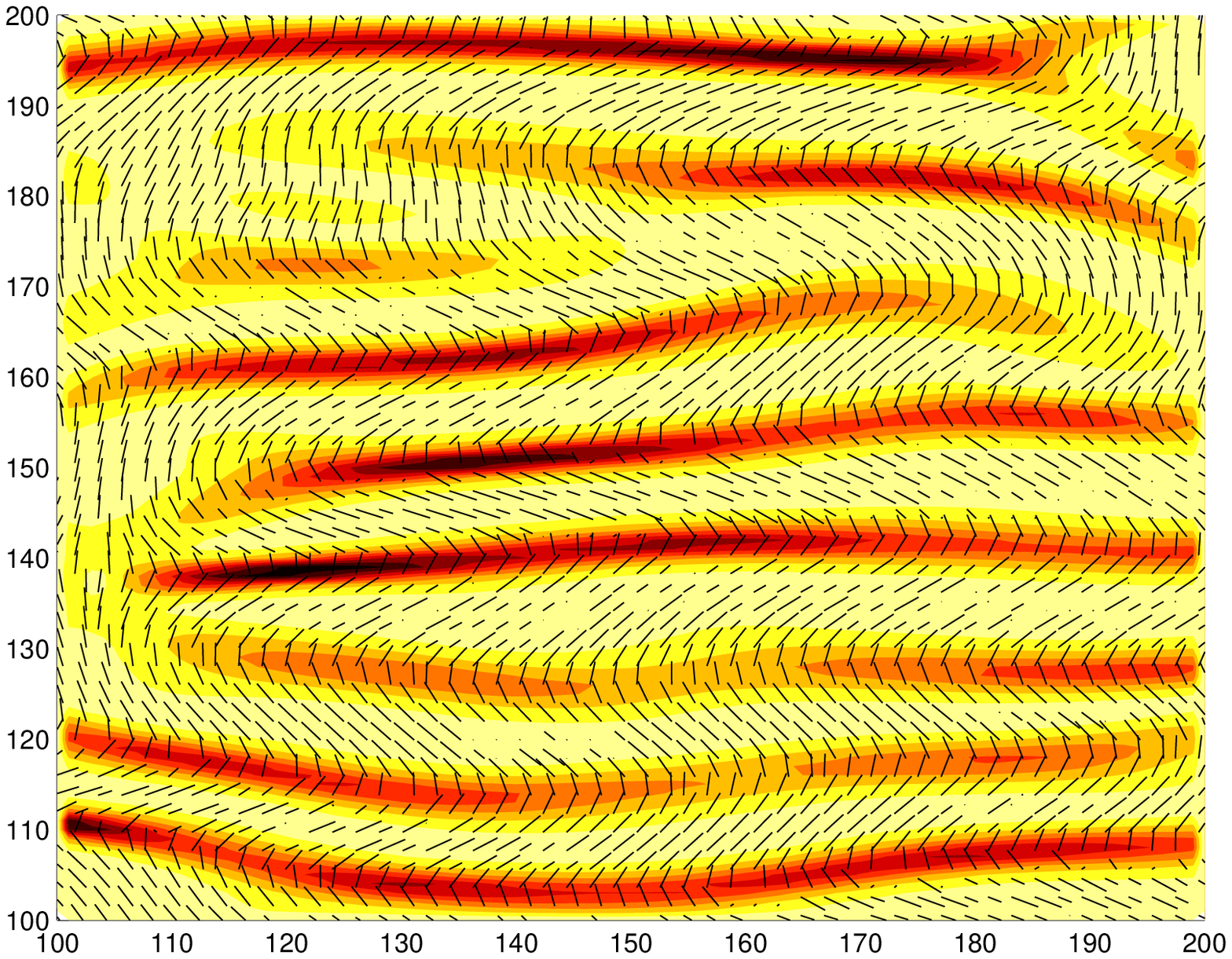}\label{fig:startupzetac}}
\vspace{10pt}
\end{minipage}
\begin{minipage}{0.59\linewidth}
\subfigure[]{\includegraphics[trim = 0 0 0 0, clip, width=\linewidth]{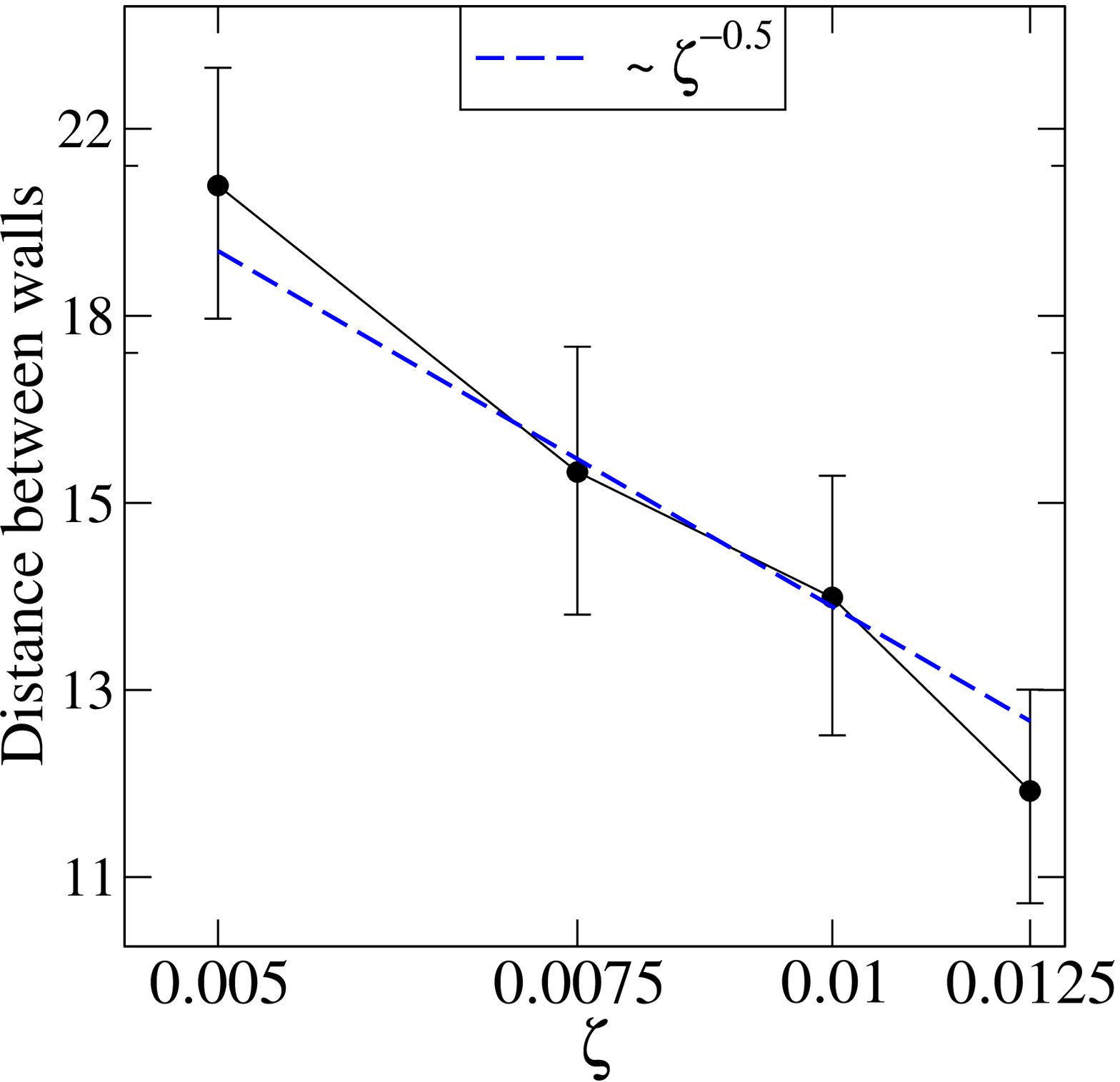}\label{fig:startupzetascale}}
\vspace{10pt}
\end{minipage}
\begin{minipage}{0.39\linewidth}
\centering
\subfigure[\ $K=0.004$]{\includegraphics[trim = 70 140 100 30, clip, width=\linewidth]{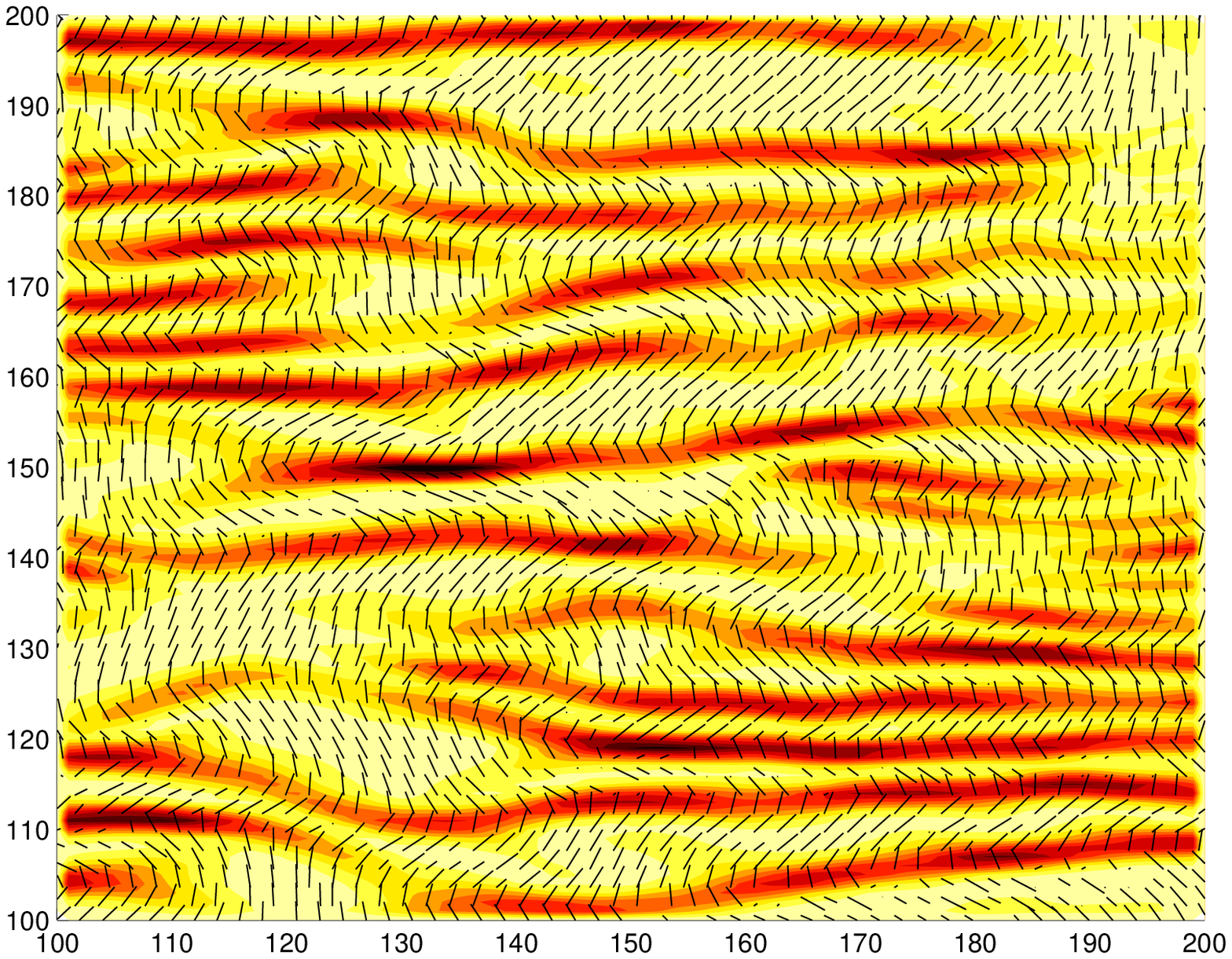}\label{fig:startupKc}}
\subfigure[\ $K=0.04$]{\includegraphics[trim = 70 140 100 30, clip, width=\linewidth]{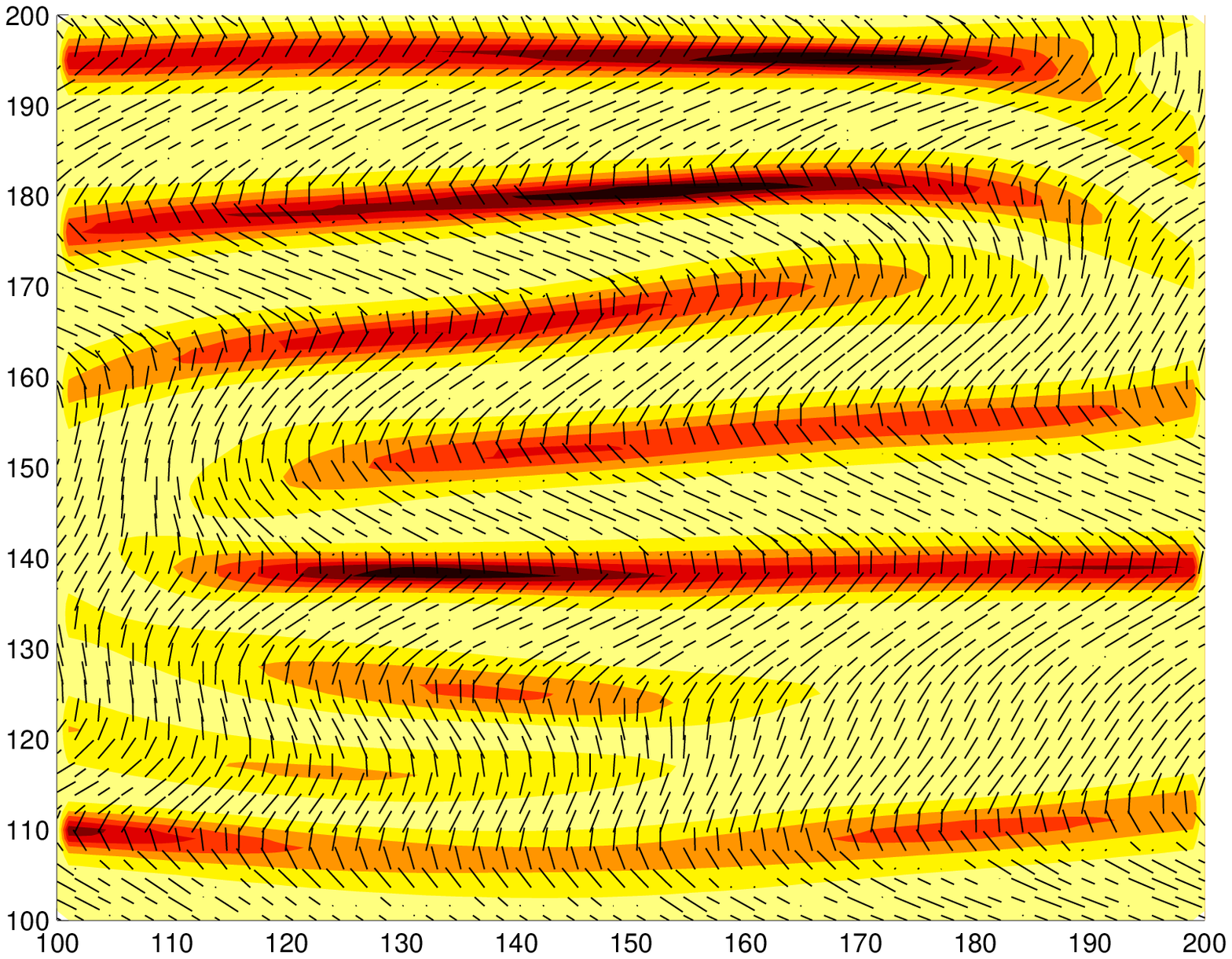}\label{fig:startupKa}}
\end{minipage}
\begin{minipage}{0.59\linewidth}
\subfigure[]{\includegraphics[trim = 0 0 0 0, clip, width=\linewidth,]{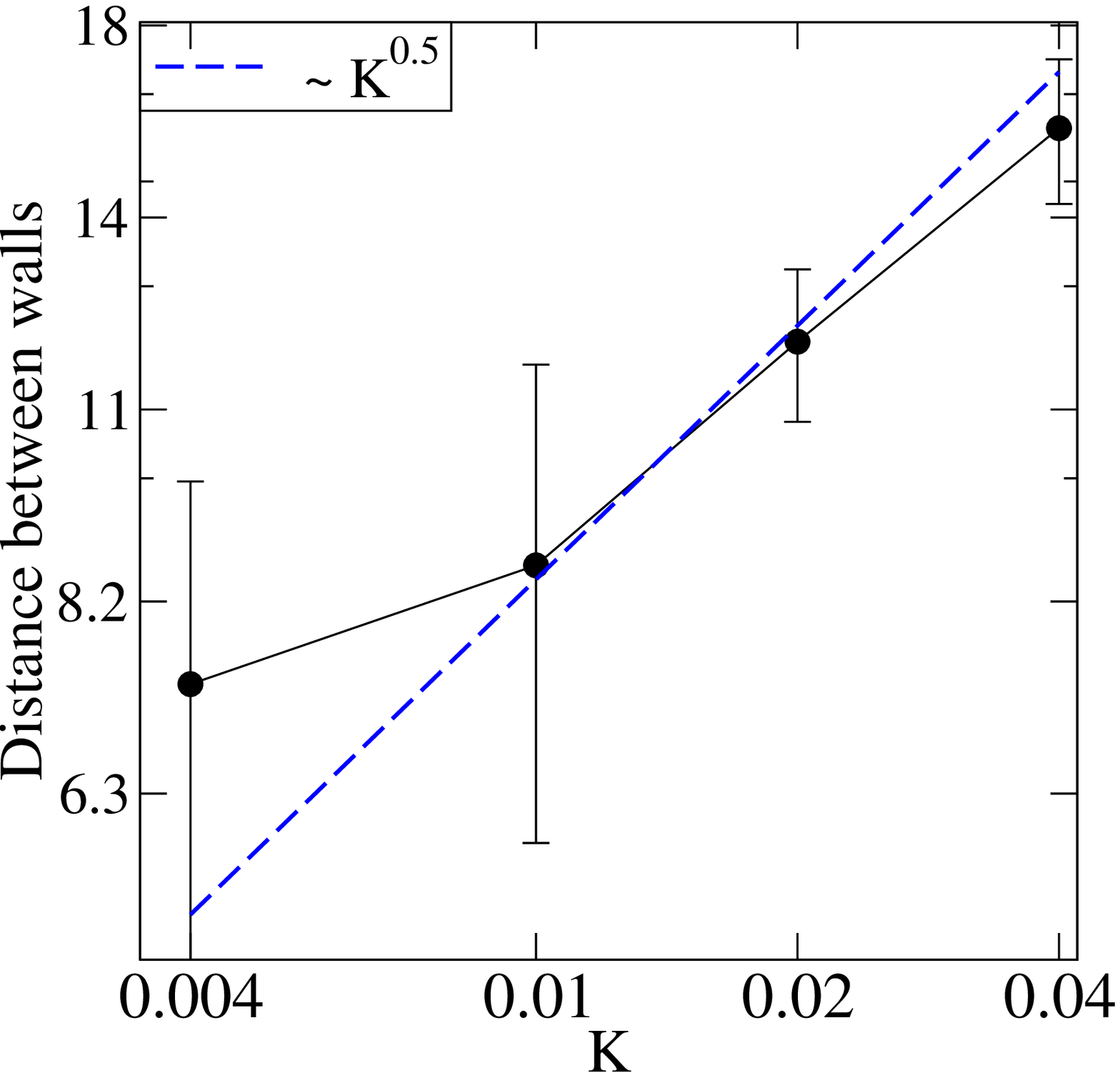}\label{fig:startupKscale}}
\end{minipage}
\caption{Wall formation resulting from hydrodynamic instabilities in an active, extensile nematic. (a)-(c) varying the activity $\zeta$; (d)-(f) varying the elastic constant $K$. The distance between walls  $\sim \sqrt{K/\zeta}$.}
\label{fig:startupzetaK}
\end{figure}

{\em Wall formation}:  The wall formation can be thought of as a direct competition between activity trying to establish a flow field by distorting the director field and elasticity trying to prevent this deformation. This competition leads to a dominant length scale $\sqrt{K/\zeta}$ for the instability \cite{Sriram2010}. We study the wall formation process in active nematics as a function of the parameters $K$ and $\zeta$. The results are illustrated in fig.~\ref{fig:startupzetaK}.

In each of these simulations, the director field was recorded just before the onset of defect pair generation. The extent of the director deformation at any point was quantified by calculating the total of the angular deviation of the director at that point from its  neighbours. This quantity will be zero in perfectly ordered regions, while it will be nonzero and large in the regions of bends. It is thus possible to identify walls quantitatively as shown in fig.~\ref{fig:startupzetaK}. Note that it is visually apparent that the distance between the walls decreases as $\zeta$ increases and as $K$ decreases as expected. To obtain a quantitative measure of the dominant length scale we take a 1D Fourier transform in the direction  normal to the walls. The resulting length scale is plotted as a function of $\zeta$ and $K$ in figs.~\ref{fig:startupzetascale} and \ref{fig:startupKscale} confirming a characteristic length scale separating the walls $\sim \sqrt{K/\zeta}$. This is strong evidence that wall formation in active nematics is indeed a consequence of the inherent hydrodynamic instability.

\begin{figure}
\subfigure[]{\subfigimgloc[trim = 110 40 80 30, clip, width=0.49\linewidth]{$t_1$}{8}{1}{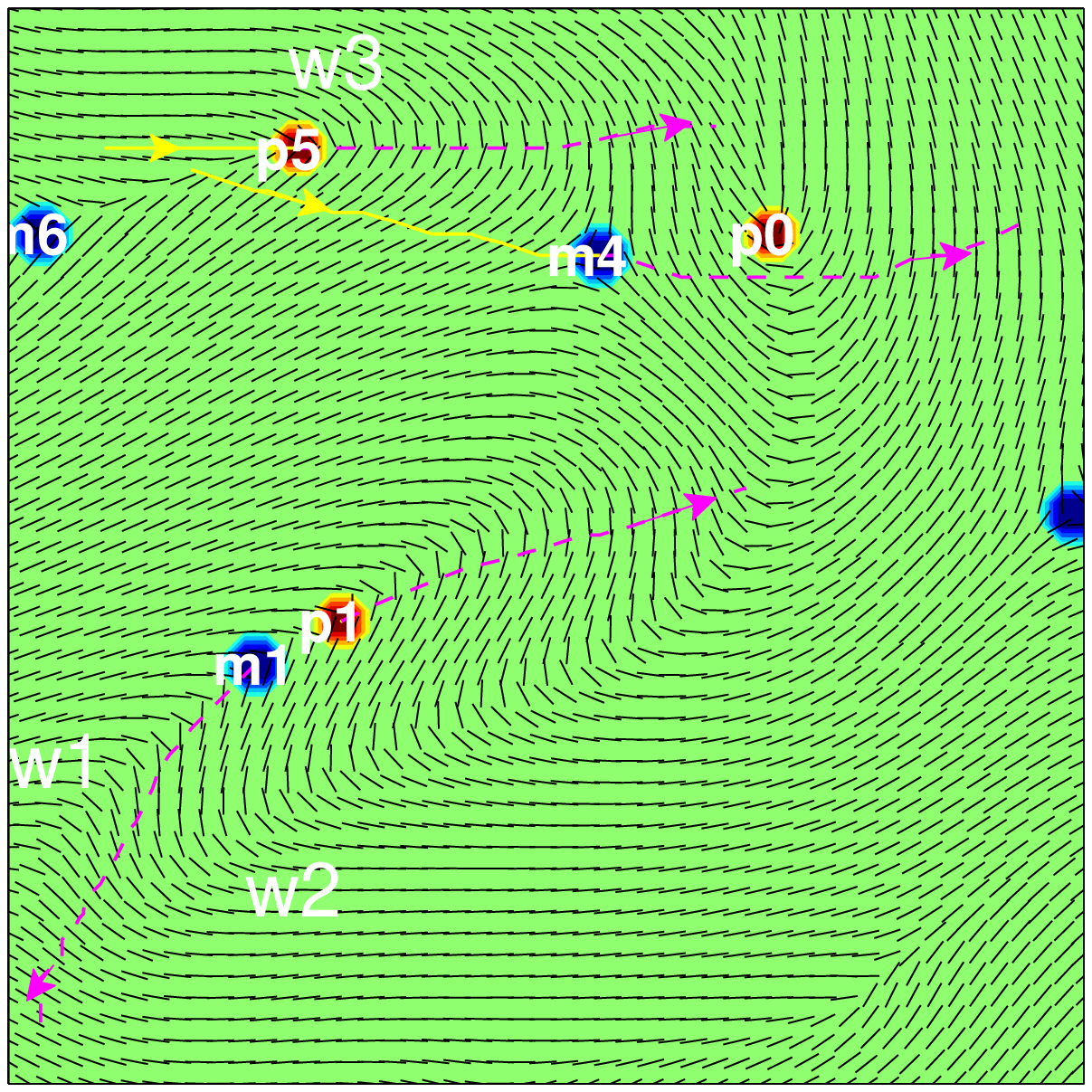}\label{fig:defdyn1}}
\subfigure[]{\subfigimgloc[trim = 110 40 80 30, clip, width=0.49\linewidth]{$t_2$}{8}{1}{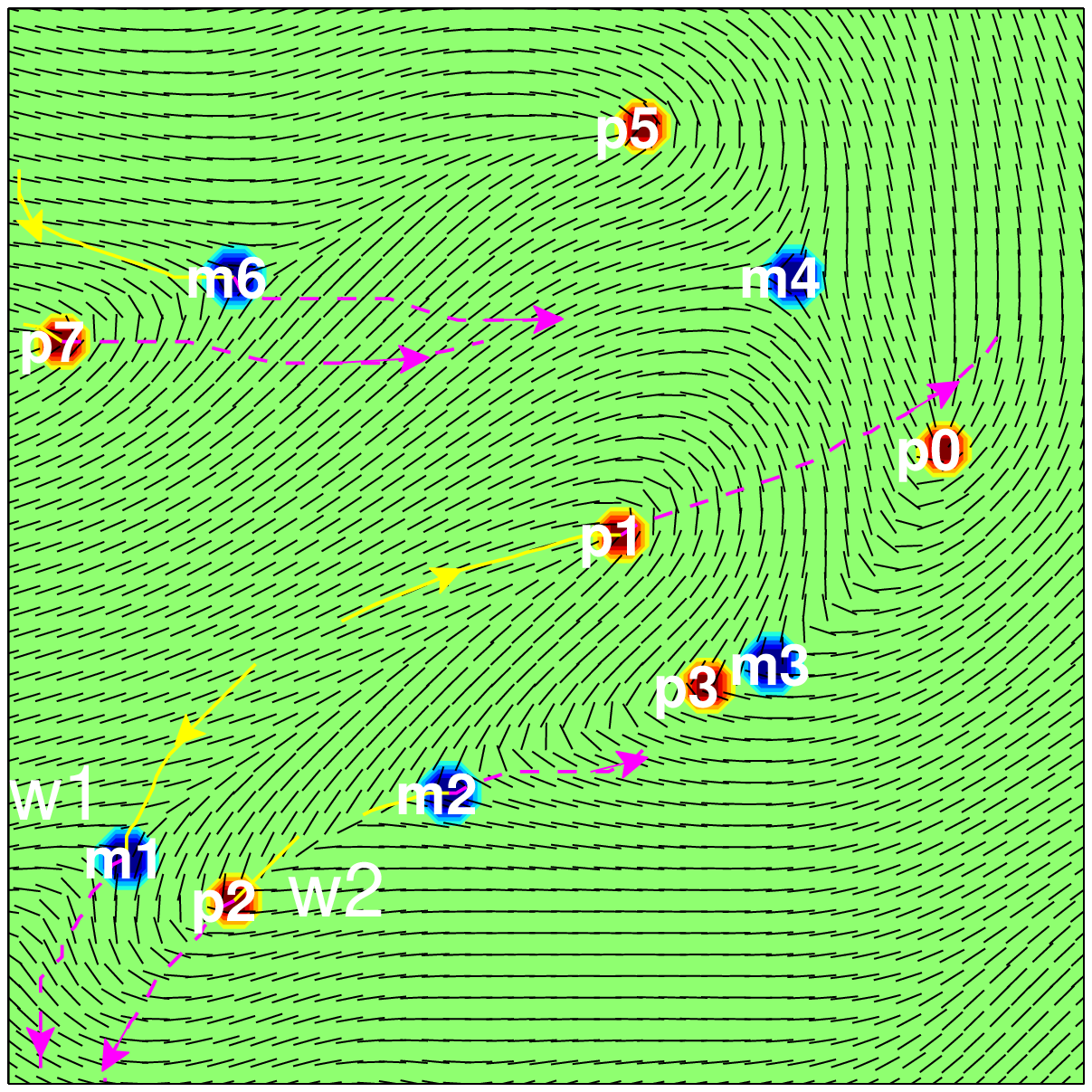}}\\
\subfigure[]{\subfigimgloc[trim = 110 40 80 30, clip, width=0.49\linewidth]{$t_3$}{8}{1}{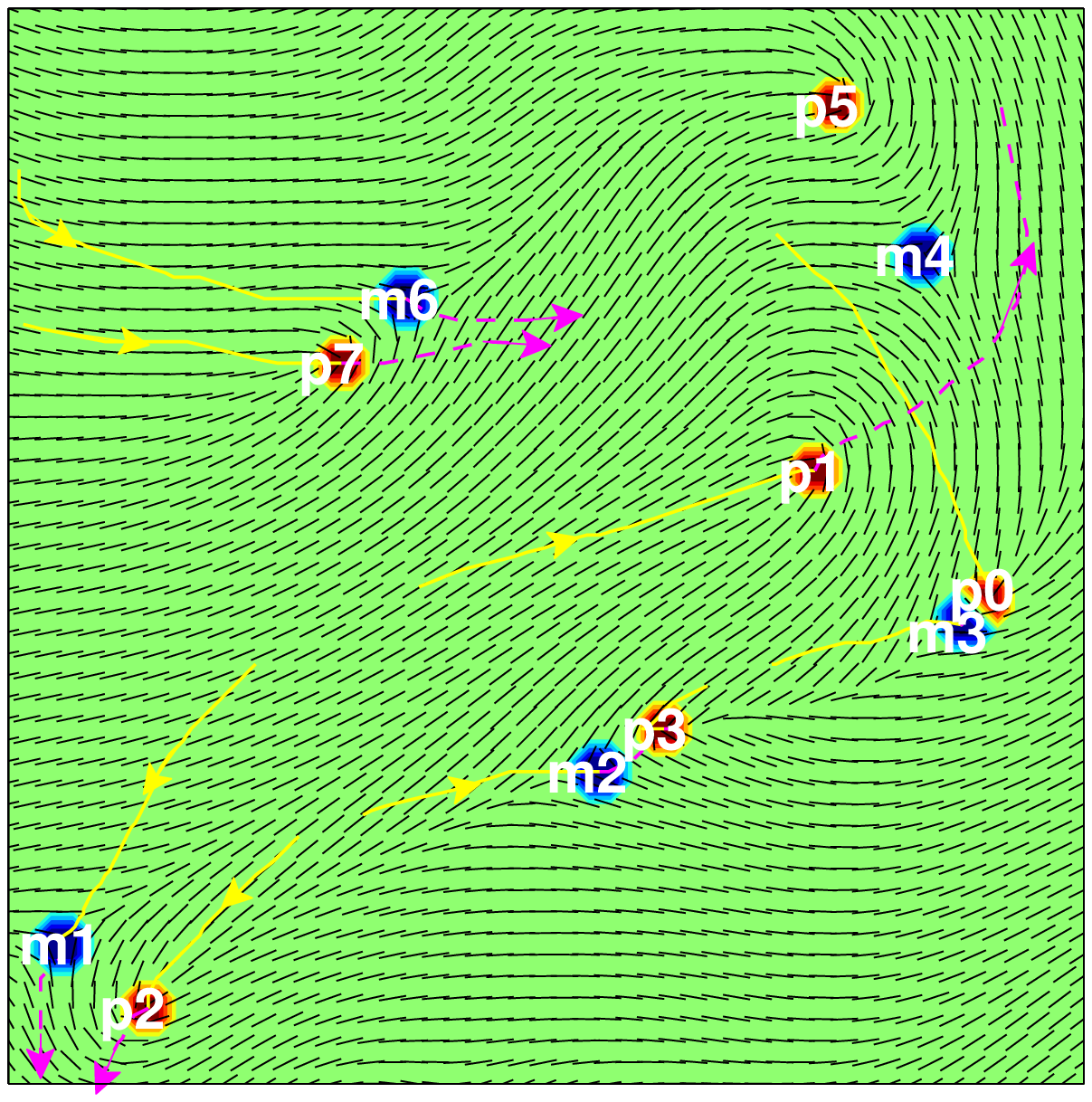}}
\subfigure[]{\subfigimgloc[trim = 110 40 80 30, clip, width=0.49\linewidth]{$t_4$}{8}{1}{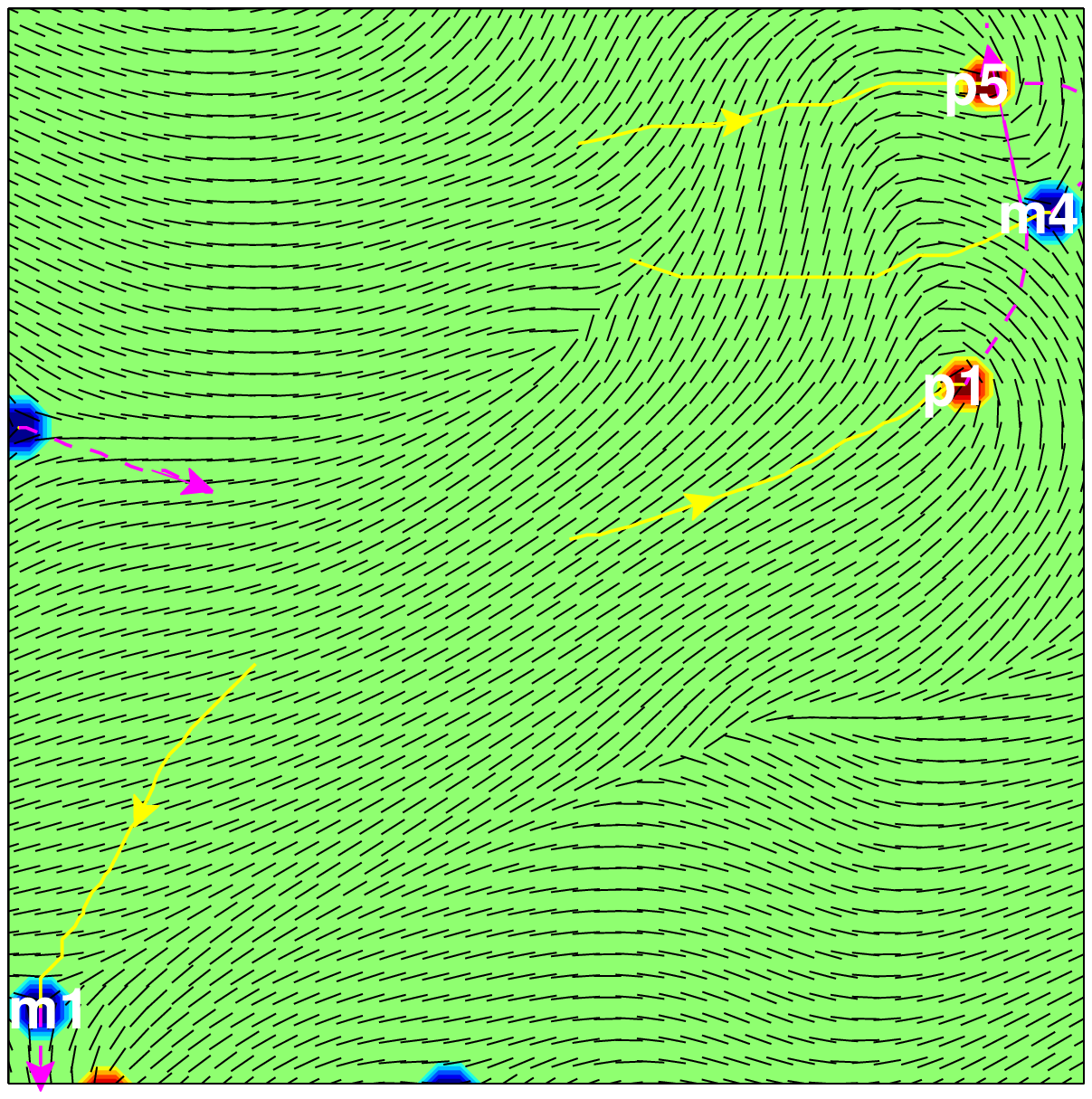}}\\
\subfigure[creation]{\includegraphics[trim = 0 0 0 0, clip, width=0.55\linewidth]{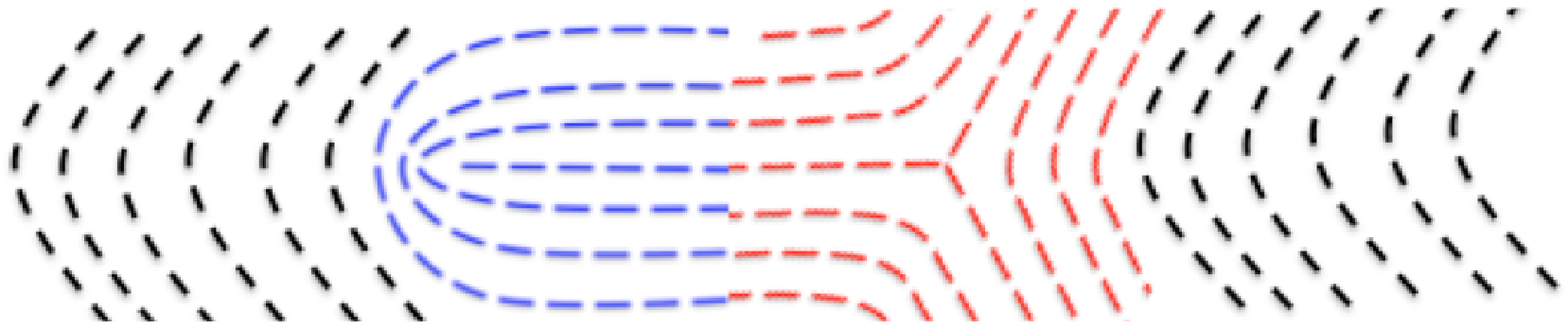}\label{fig:defcre}}
\subfigure[annihilation]{\includegraphics[trim = 0 0 0 0, clip, width=0.41\linewidth]{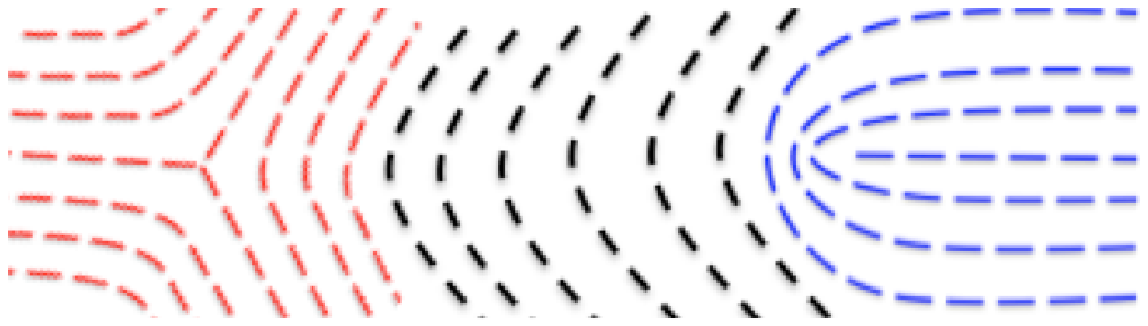}\label{fig:defann}}
\caption{Snapshots of defects in an extensile suspension at successive times., $t_1 \ldots t_4$.  Walls are labelled as {\it w1}, {\it w2} $\ldots$, positive defects by {\it p1}, {\it p2} $\ldots$, and negative defects by {\it m1}, {\it m2} $\ldots$.
The past and future trajectories of the defects are shown with continuous (yellow) and dashed (magenta) lines respectively. There are several cases of defect formation, eg {\it m1}--{\it p1} in (a),  {\it m2}--{\it p2} and  {\it m3}--{\it p3} in (b). 
The defects move along the walls restoring them to a nematic configuration. Annihilation events also occur eg  {\it m2}--{\it p3},  {\it m3}--{\it p0},  {\it m6}--{\it p7} in (c). (e),(f): schematic illustration of defect pair creation and annihilation. The relative orientation of defects in these events are different, as is also evident in panels (a)-(d).}
\label{fig:defformn}
\end{figure}

{\em Defect formation and annihilation:} Domain walls are known to be unstable in passive liquid crystals, giving rise to a pair of defects. However, the  literature indicates that this happens due to a difference in elastic constants and  known mechanisms are in three dimensions \cite{Rey2002,Lozar2005}. In active nematics, a local perturbation of the wall can nucleate pairs of oppositely charged  $\pm 1/2$ defects. Such fluctuations are enhanced by the vortical flow field. Before we analyse the details of defect formation mechanisms, we describe an overall picture of the defect dynamics.

Snapshots of the nucleation process, and the subsequent defect motion leading to annihilation events, are shown in fig.~\ref{fig:defformn}. At time $t_1$, three different walls, labelled as {\it w1}, {\it w2} and {\it w3} can be identified. Nucleation of a pair of defects {\it m1-p1} from {\it w1} occurs at $t=t_1$. Two other pairs, {\it m2-p2} and {\it m3-p3} are formed in wall {\it w2} at time $t=t_2$. Arrows in the figure show the trajectories of the defects. When the defects form they have a propensity to move along the walls, driven by elastic forces and flow. This causes the walls to `unzip' i.e, to relax back to the nematic state. For example,  {\it w1} is unzipped by the defect pair {\it m1-p1} during $t_1 \le t \le t_2$ and {\it w2} by both {\it m2-p2} and {\it m3-p3} during  $t_2 \le t \le t_3$. A defect {\it p5} unzipping wall {\it w3} for $t \ge t_1$ is also visible.

When a defect encounters an oppositely charged defect they annihilate each other. For example,  {\it m2} and {\it p3}, {\it m3} and {\it p0}, {\it m6} and {\it p7} meet at $t \approx t_3$ resulting in the annihilation of each of these pairs. Formation of more than one closely spaced pair in a wall (e.g.~{\it m2-p2} and {\it m3-p3}) tends to result in fast annihilation as the defects move easily along the walls (e.g.~{\it m2} with {\it p3}), and the wall disappears. As the activity increases the defects are more likely to be driven  away from the wall in which they have formed by the ambient flow. This tends to lead to longer times between creation and finding an oppositely charged defect with the correct orientation for annihilation (e.g.~{\it m6} and {\it p7}). Indeed in general the route to defect annihilation depends on the local order and flows.  For example, {\it m4} is weakly associated with two different walls {\it w1} and {\it w3} at $t=t_1$ and hence there is the  possibility of {\it m4} annihilating with {\it p5} or {\it p1}.  As time proceeds {\it m4}--{\it p5} annihilation occurs ($t>t_4$, not shown) while {\it p1} moves away from the original wall. 

One might ask why oppositely charged defects do not annihilate each other immediately after they form\cite{Giomi2013}. This is  because creation and annihilation occur for different orientations of defects as illustrated in fig.~\ref{fig:defcre}-\ref{fig:defann}.
Fig.~\ref{fig:defcre} shows that, as the two defects move apart, the length of wall between them  regains nematic order (horizontal alignment in fig.~\ref{fig:defcre}). Thus the process of defect pair creation is a very natural way to relieve the bending energy of the wall. Similarly fig.~\ref{fig:defann} shows that, as defects annihilate, the stretch of wall between them is removed.
Realigned nematic regions then undergo further hydrodynamic instabilities and the system reaches a dynamical steady state.  Similar defect creation and annihilation events are observed in experiments on microtubule bundles driven by kinesin molecular motors \cite{Dogic2012}.

\begin{figure}
\centering
\subfigure{\subfigimg[trim = 0 0 0 0, clip, width=0.9\linewidth]{(a)}{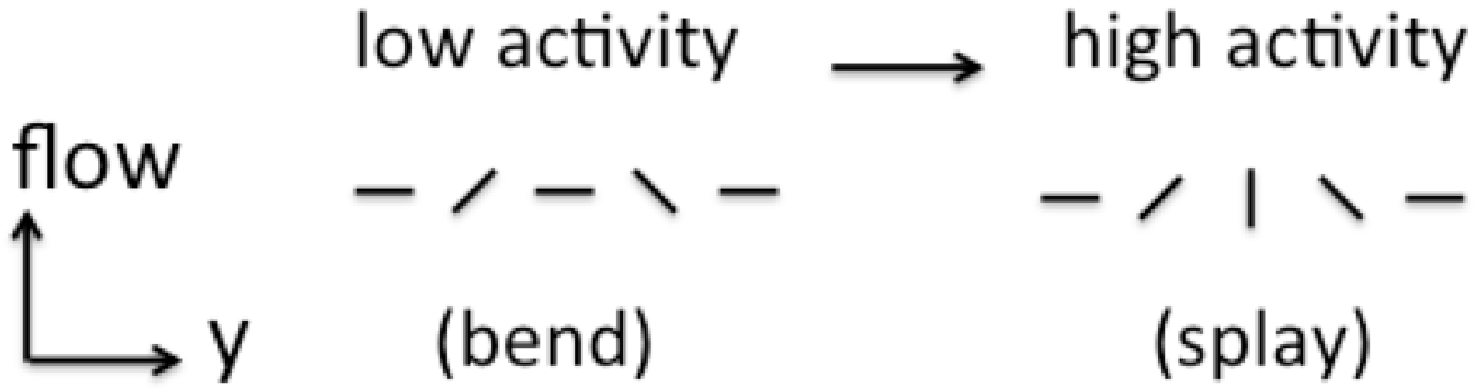}\label{fig:defect1d}}
\subfigure{\subfigimgloc[trim = 10 5 0 10, clip, width=0.49\linewidth]{\colorbox{white}{(b)}}{17}{1.1}{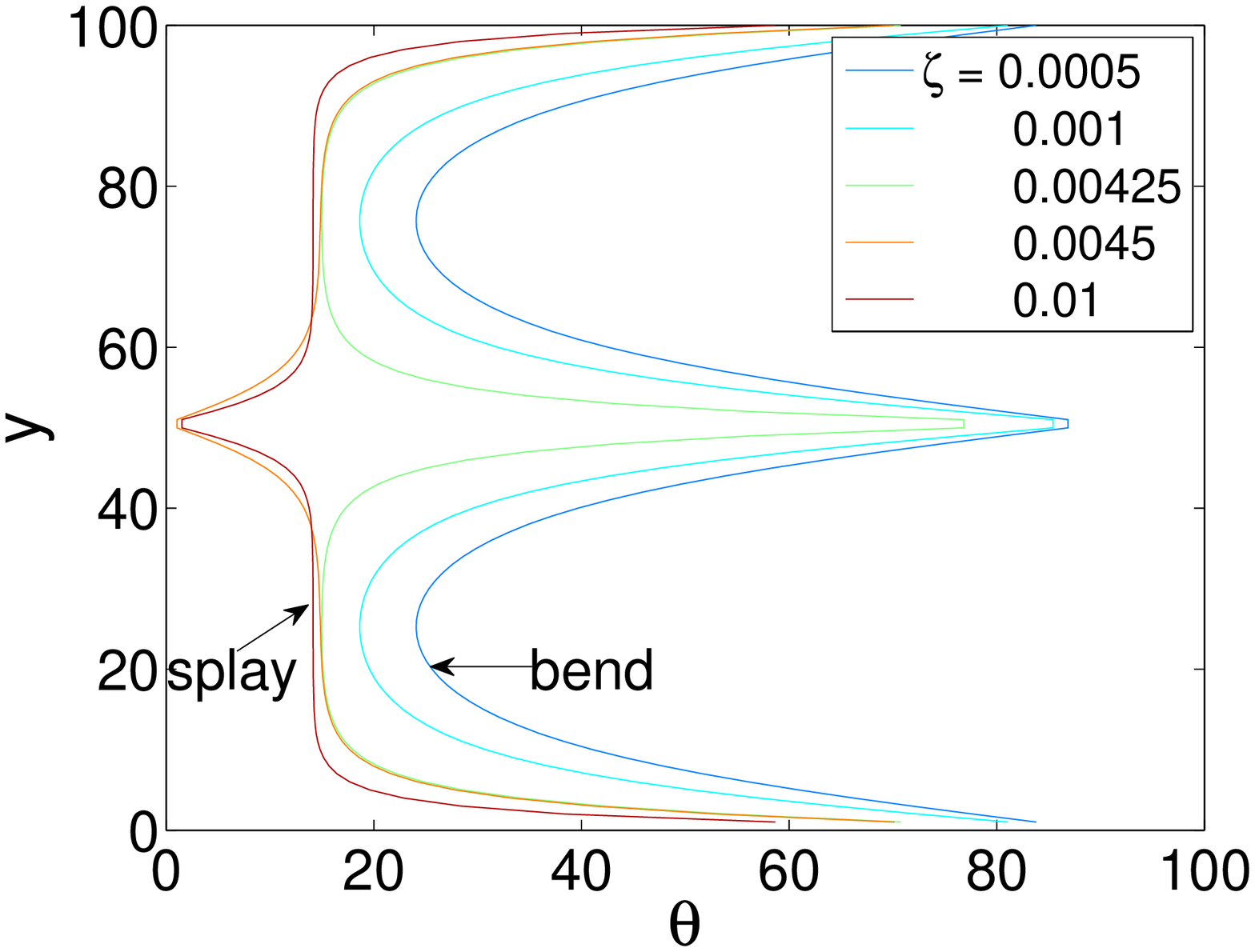}\label{fig:chanactq}}
\subfigure{\subfigimgloc[trim = 10 5 0 10, clip, width=0.49\linewidth]{\colorbox{white}{(c)}}{17}{1.1}{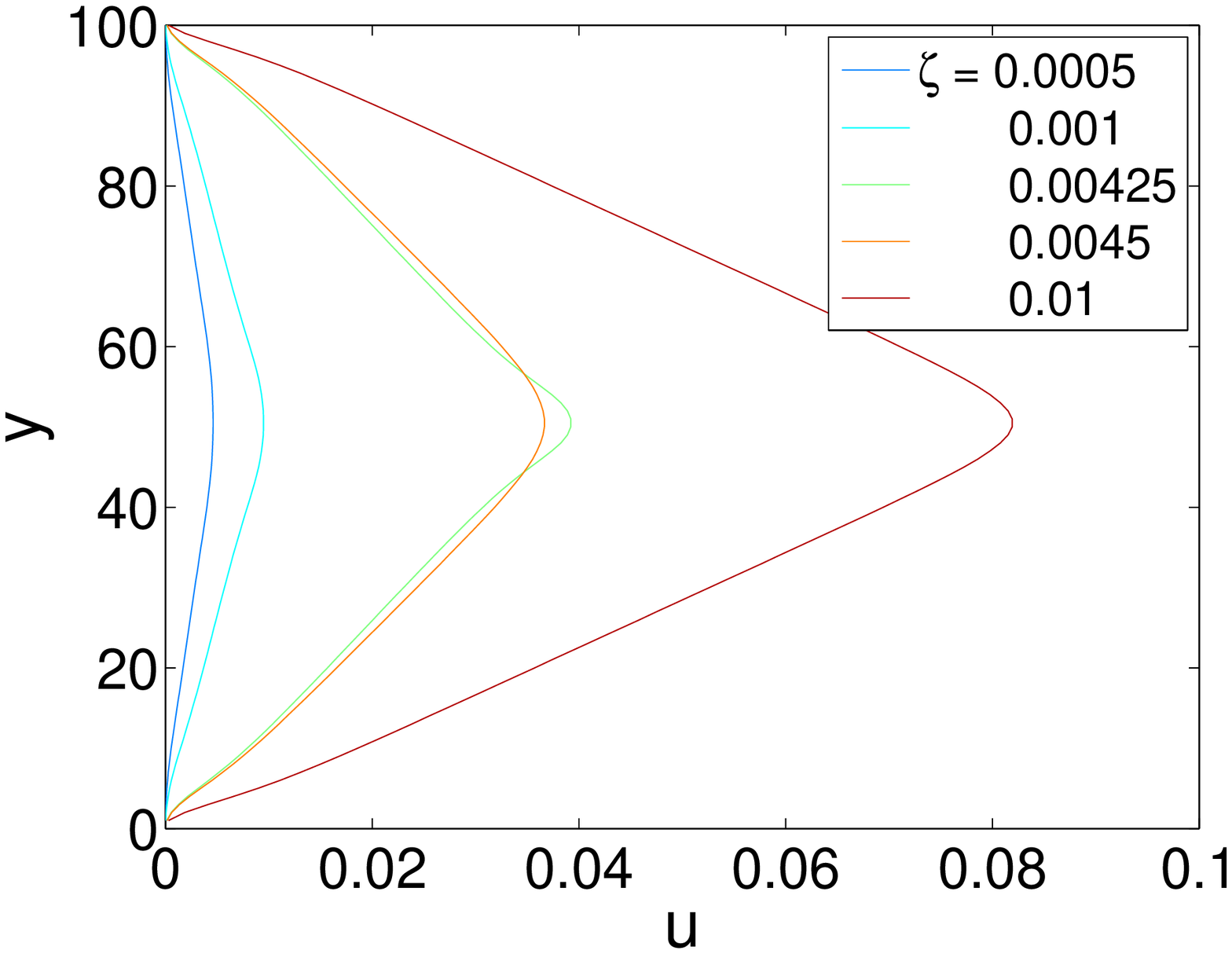}\label{fig:chanactu}}
\caption{(a) Bend to splay transition of the director field for increasing $\zeta$ in a channel with homeotropic boundary conditions. This is similar to the director change at the point of nucleation of a defect pair in a wall. (b) Variation of the director profile across the channel with activity $\zeta$.  (c) Corresponding velocity profiles.}
\label{fig:chanact}
\end{figure}

{\em Flow-driven defect formation}: 
We next consider in more detail how the topological defects are created. As in passive systems their formation is driven by  minimisation of elastic free energy, but in active systems, this is not the only driving force. Activity, which generates flow, can also play an important role. To illustrate this we consider,
as a model system, a channel with any gradients only in the direction normal to the boundary. The spontaneous symmetry breaking which results in a unidirectional steady state flow in these channels above a threshold activity is well established \cite{Joanny2005, Davide2007}. For homeotropic anchoring of the director field on the boundaries the instability is to a state where the director field adopts a bend configuration similar to one of the active walls in fig.~\ref{fig:startupzetaK}. As activity increases, the shear increases, and the bend configuration becomes unstable to a splay configuration as shown in fig.~\ref{fig:defect1d}.  In figs.~\ref{fig:chanactq} and \ref{fig:chanactu} we show the corresponding changes in the director and velocity profiles across the channel. For this simple model  the bend to splay transition occurs uniformly throughout the length of the channel.
Returning to the full 2D geometry, and unbounded walls, it is apparent that a similar bend to splay transition 
occurs at the point of nucleation of a defect pair in the wall. The difference is that because the 1D symmetry is lost due to, for example, local flows the bend to splay transition takes place at one point. The splay region then expands in both directions along the wall, corresponding to the defects moving apart. 
 
Parameters influencing the  instability of the bend configurations can be identified from an analysis of eqs.~(\ref{eqn:lc}) and (\ref{eqn:ns}) assuming 1D and an order parameter of constant magnitude. With these simplifying assumptions eq.~(\ref{eqn:ns}) reduces, in the steady state, to
\begin{align}
0 = \frac{1}{2} \frac{du}{dy} \left(\lambda_1 \cos{2\theta} -1 \right) + \Gamma K \frac{d^2\theta}{dy^2}
\label{eqn:leslieq}
\end{align}
where $y$ is the transverse direction and $\theta$ is the director angle to the unidirectional flow field, $u$. $\lambda_1 = (3q+4)\lambda/9q$ is related to the flow alignment parameter  $\lambda$ of eq.~(\ref{eqn:lc}) and $q$ is the magnitude of the nematic order, the largest eigenvalue of $\mathbf{Q}$ \cite{Davide2007}. If the active stress balances the viscous stress then, from eq.~(\ref{eqn:lc}) we also obtain
\begin{align}
\frac{du}{dy} = \frac{\zeta}{2 \mu} \sin{2\theta} .
\label{eqn:lesliens}
\end{align}
For homeotropic boundary conditions, one obtains the director profiles shown in fig.~\ref{fig:chanactq}
\cite{Davide2007}. However, $\theta_c$, the value of $ \theta$ at the centre of the channel, may be $\theta_c=0^{\circ}$ or $\theta_c=90^{\circ}$, corresponding to a bend configuration or a splay configuration respectively, depending upon the activity.
If the director profile is slightly perturbed around the steady state variations $\hat{\theta}$ of $\theta_c$ at the centre of the channel evolve as 
\begin{align}
\frac{d\hat{\theta}}{dt} &= -\frac{ \lambda_1 \zeta \hat{\theta}}{2 \mu} \sin^2{2\theta_c} \nonumber\\
&+ \frac{\zeta}{2\mu} \hat{\theta} \cos{2\theta_c} \left(\lambda_1 \cos{2\theta_c} -1 \right) + \Gamma K \frac{d^2\hat{\theta}}{dy^2} .
\label{eqn:thetacap}
\end{align}
The growth of $\hat{\theta}$ depends upon the sign of the terms on the right hand side of eq.~(\ref{eqn:thetacap}).  If  $({\zeta}/{\mu \Gamma K})$ is sufficiently large, the coupling between the flow and the director field (the first two terms) will dominate over the diffusive mechanism. For $\theta_c \approx 0^{\circ}$ and  $\theta_c \approx 90^{\circ}$
these terms reduce to $\frac{\zeta}{2\mu}\hat{\theta}(\lambda_1 - 1)$ and $\frac{\zeta}{2\mu}\hat{\theta}(\lambda_1 + 1)$ respectively. The second expression is always positive and gives rise to exponentially growing modes of $\hat{\theta}$.
Thus the coupling of the director field to the flow will tend to drive a bend to splay transition at high activities, in agreement with the simulations.

 \begin{figure}
\begin{minipage}{0.48\linewidth}
\subfigure[]{\includegraphics[trim = 70 140 50 130, clip, width=\linewidth]{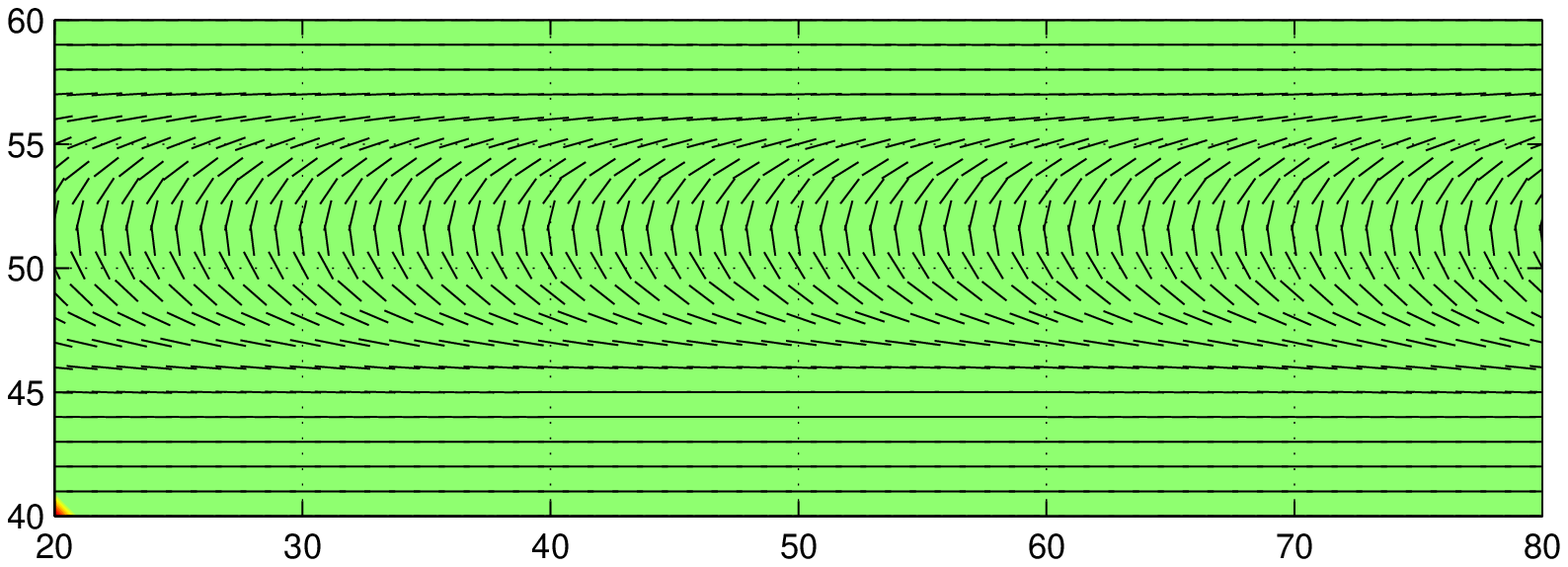}\label{fig:2ddef0}}
\subfigure[]{\includegraphics[trim = 250 140 150 130, clip, width=0.38\linewidth]{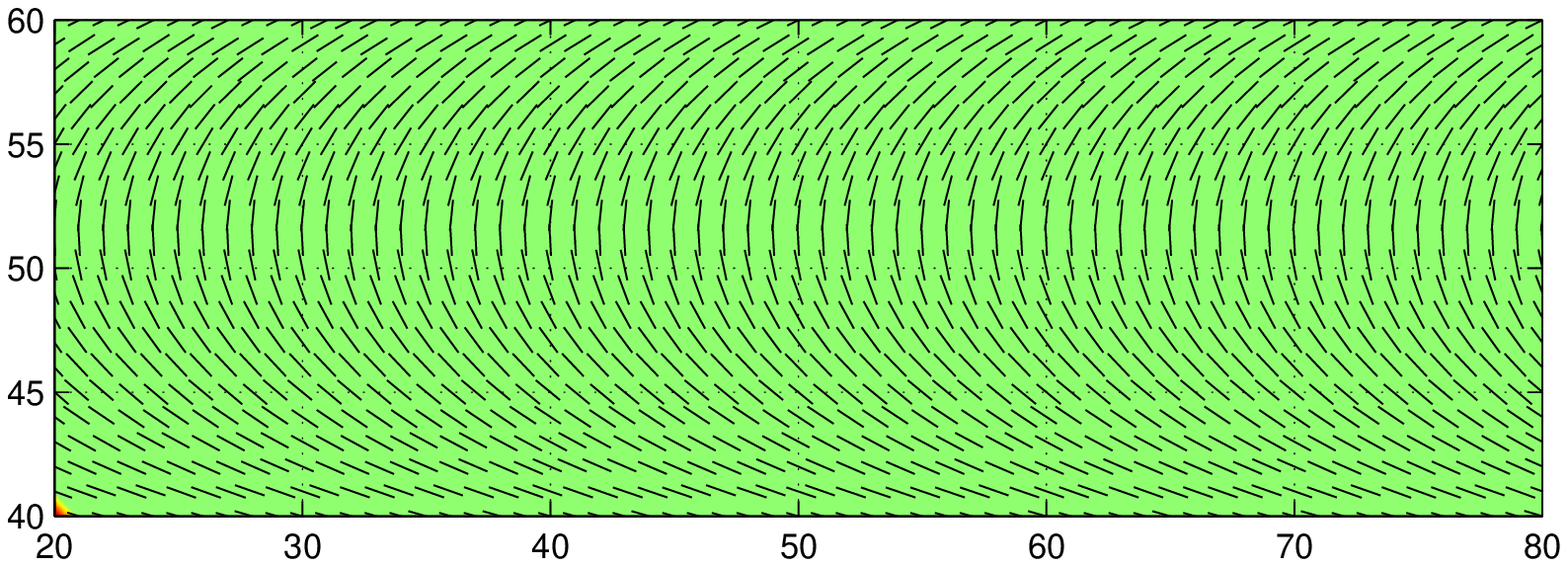}\label{fig:2dsimp}}
\subfigure[]{\includegraphics[trim = 165 140 150 130, clip, width=0.58\linewidth]{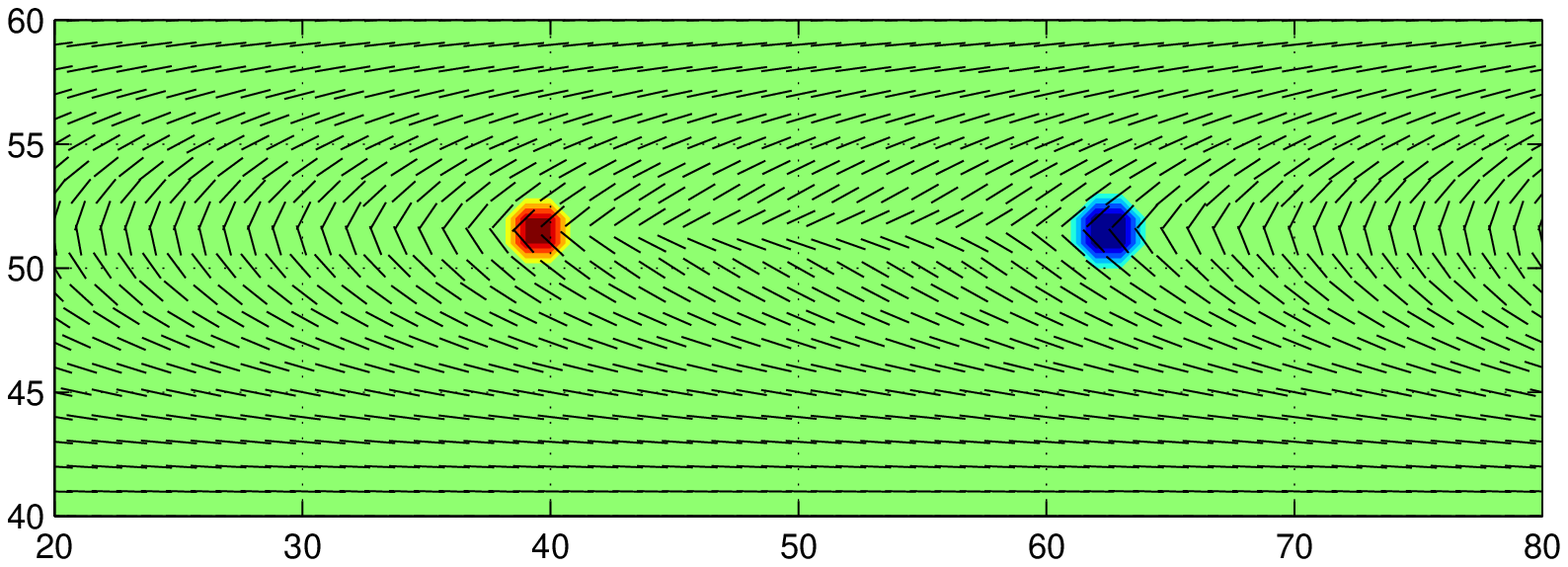}\label{fig:2ddef1}}
\end{minipage}
\begin{minipage}{0.48\linewidth}
\subfigure[]{\includegraphics[trim = 0 0 0 0, clip, width=\linewidth]{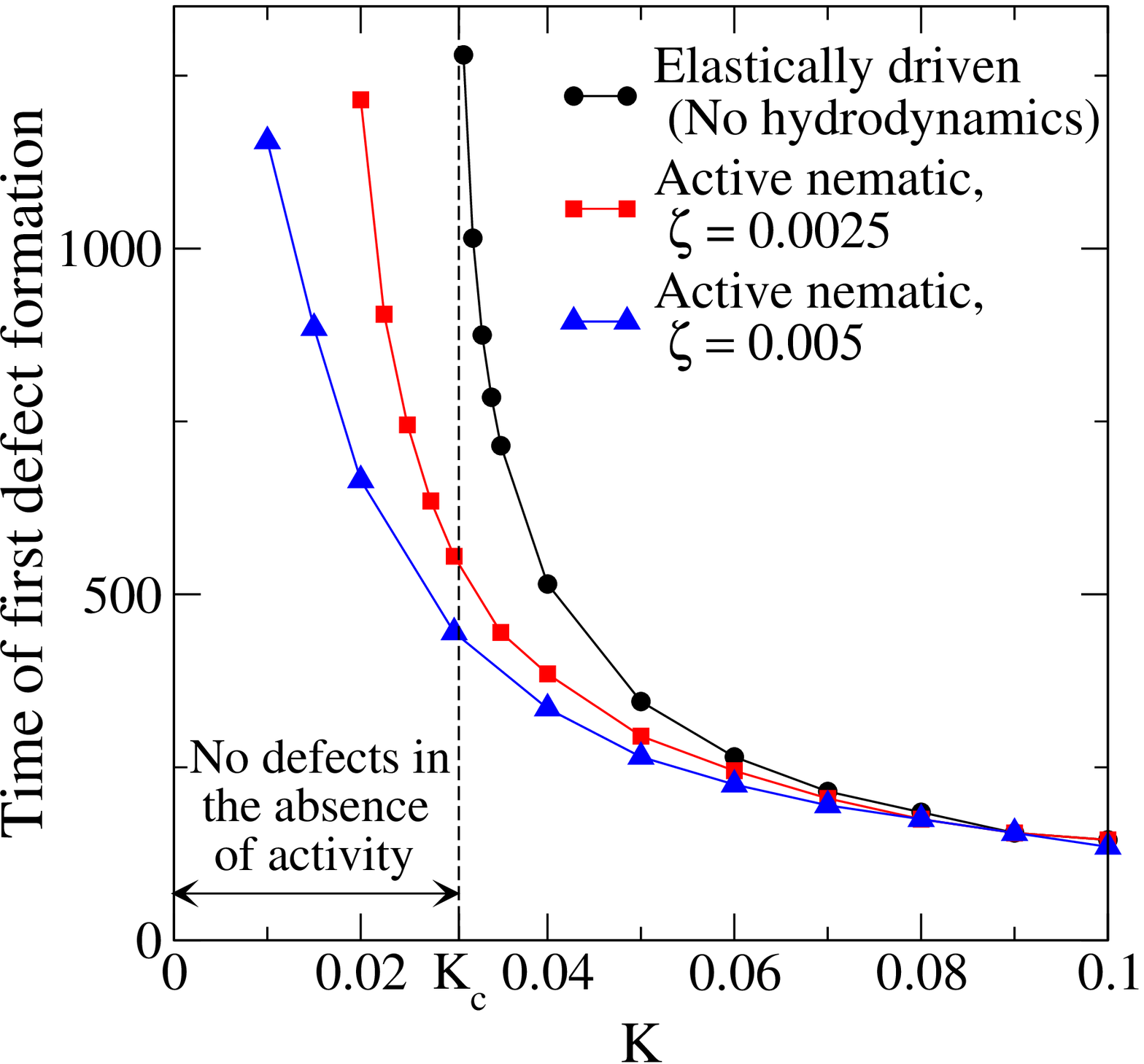}\label{fig:defKdep}}
\end{minipage}
\caption{(a) Initial conditions leading to elastically driven (b) simple relaxation and (c) defect formation in a wall. (d) Time required for the formation of the first pair of defects  as a function of the elastic constant $K$ and the activity $\zeta$.}
\label{fig:2ddef}
\end{figure}

 {\em Elastic defect formation}: 
Flow is, however, not essential to create defects. To demonstrate that defect pair formation can be driven purely by elastic energy we solved  eq.~(\ref{eqn:lc}) with $\mathbf{u}=0$. 
Using as initial condition a wall in a nematic domain (fig.~\ref{fig:2ddef0}) the wall either relaxed continuously to a vertically aligned nematic (fig.~\ref{fig:2dsimp}) or a pair of defects were created (fig.~\ref{fig:2ddef1}) leading to a horizontally aligned nematic depending upon K.
The defect formation is faster when $K$ increases as illustrated in fig.~\ref{fig:defKdep}. This figure also shows the balance between the elasticity-driven and flow-driven defect formation processes.  For example, for the particular configuration and parameters used, there is a critical $K_c$ below which no defects form for $\zeta=0$. However non-zero activity allows defects to form even below $K_c$. For $K>>K_c$ the time to defect formation is unaffected by the active flow.



\begin{figure}[b]
\subfigure{\subfigimgloctwice[trim = 120 45 95 75, clip, width=0.48\linewidth]{\wb{{(a)}}}{3}{1}{$t_2$}{100}{1}{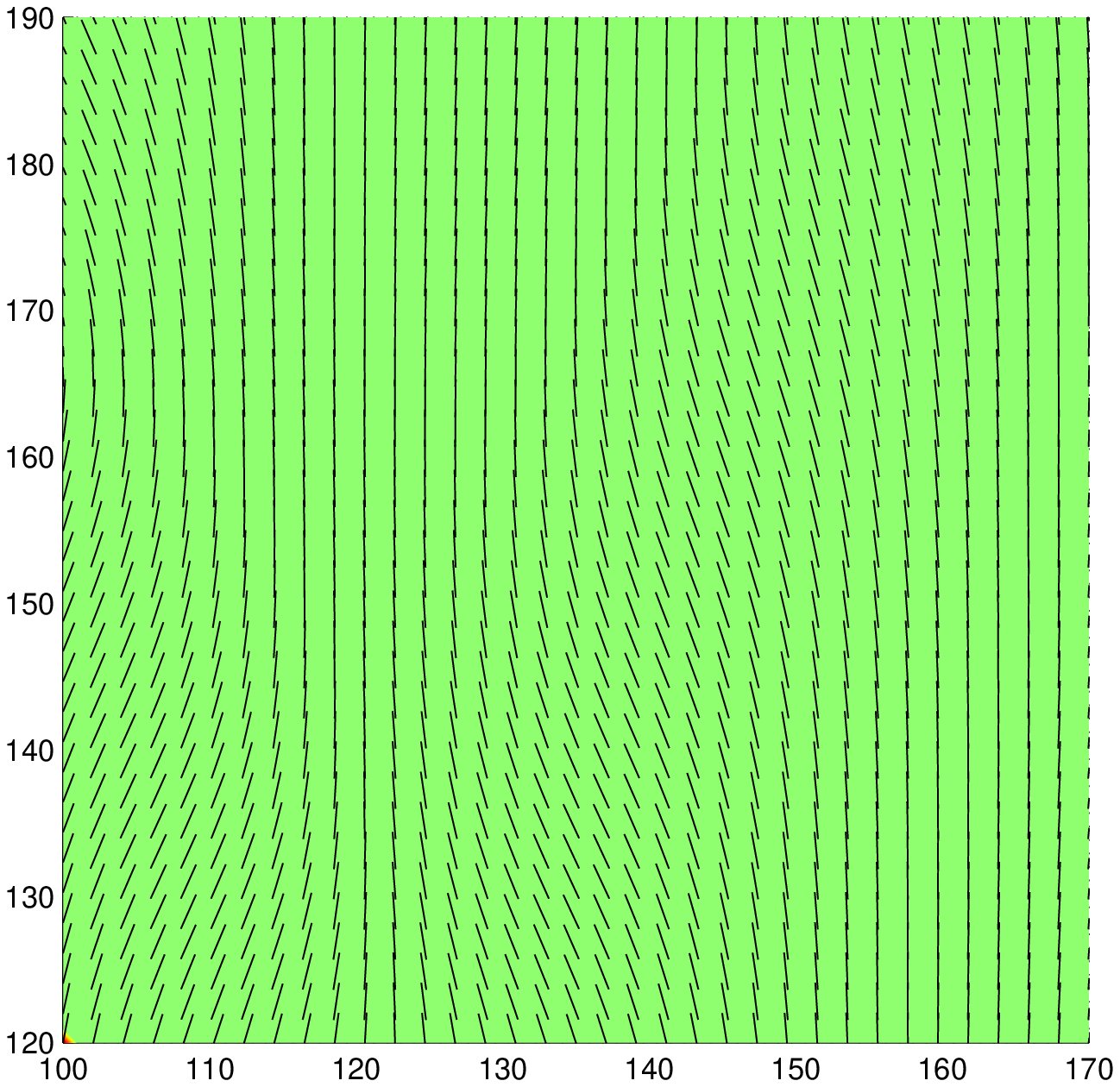}\label{fig:startupcont2}}
\subfigure{\subfigimgloctwice[trim = 120 45 95 75, clip, width=0.48\linewidth]{\wb{{(b)}}}{3}{1}{$t_3$}{100}{1}{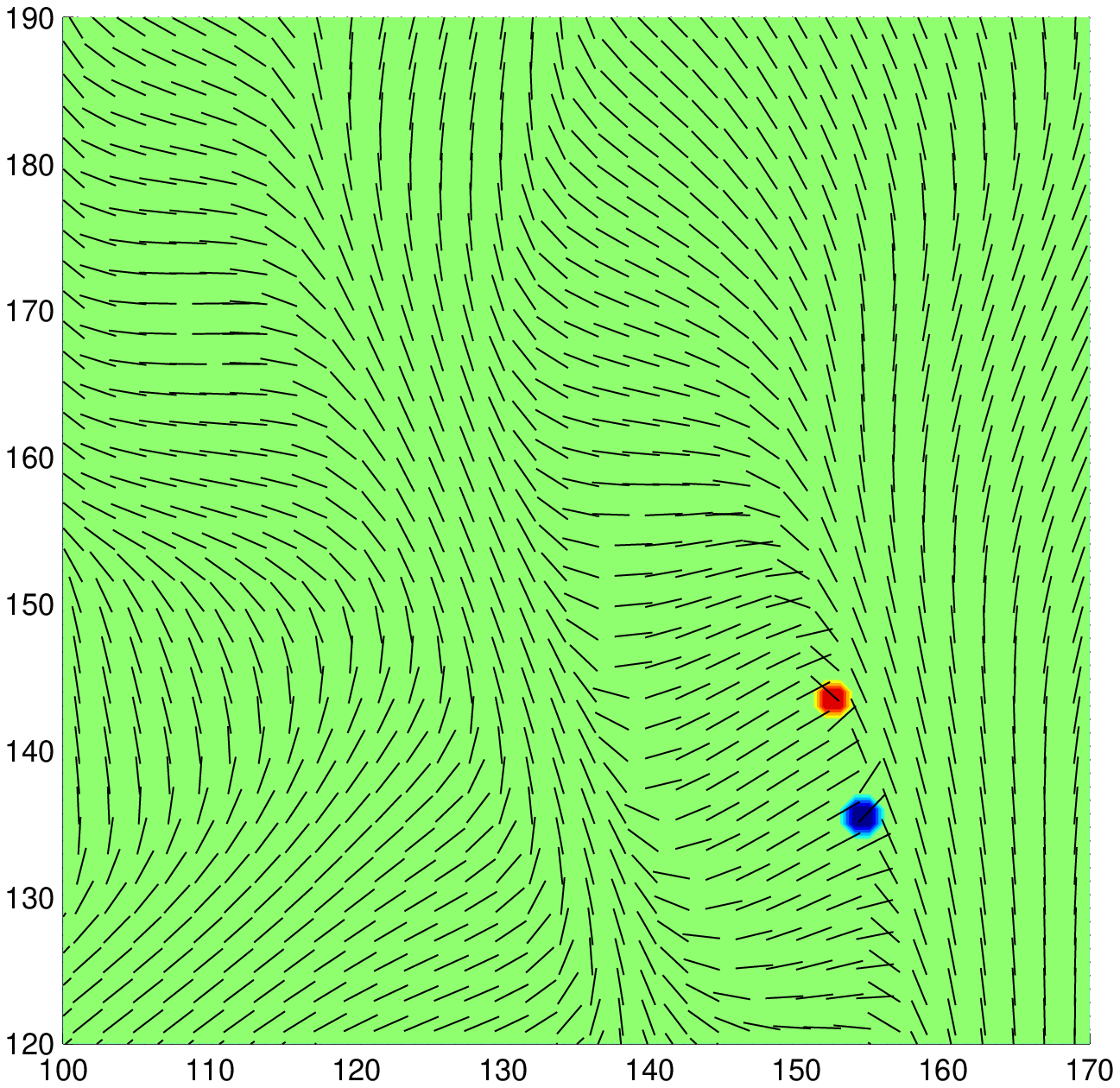}\label{fig:startupcont3}}\\
\subfigure{\subfigimgloctwice[trim = 120 45 95 75, clip, width=0.48\linewidth]{\wb{{(c)}}}{3}{1}{$t_4$}{100}{1}{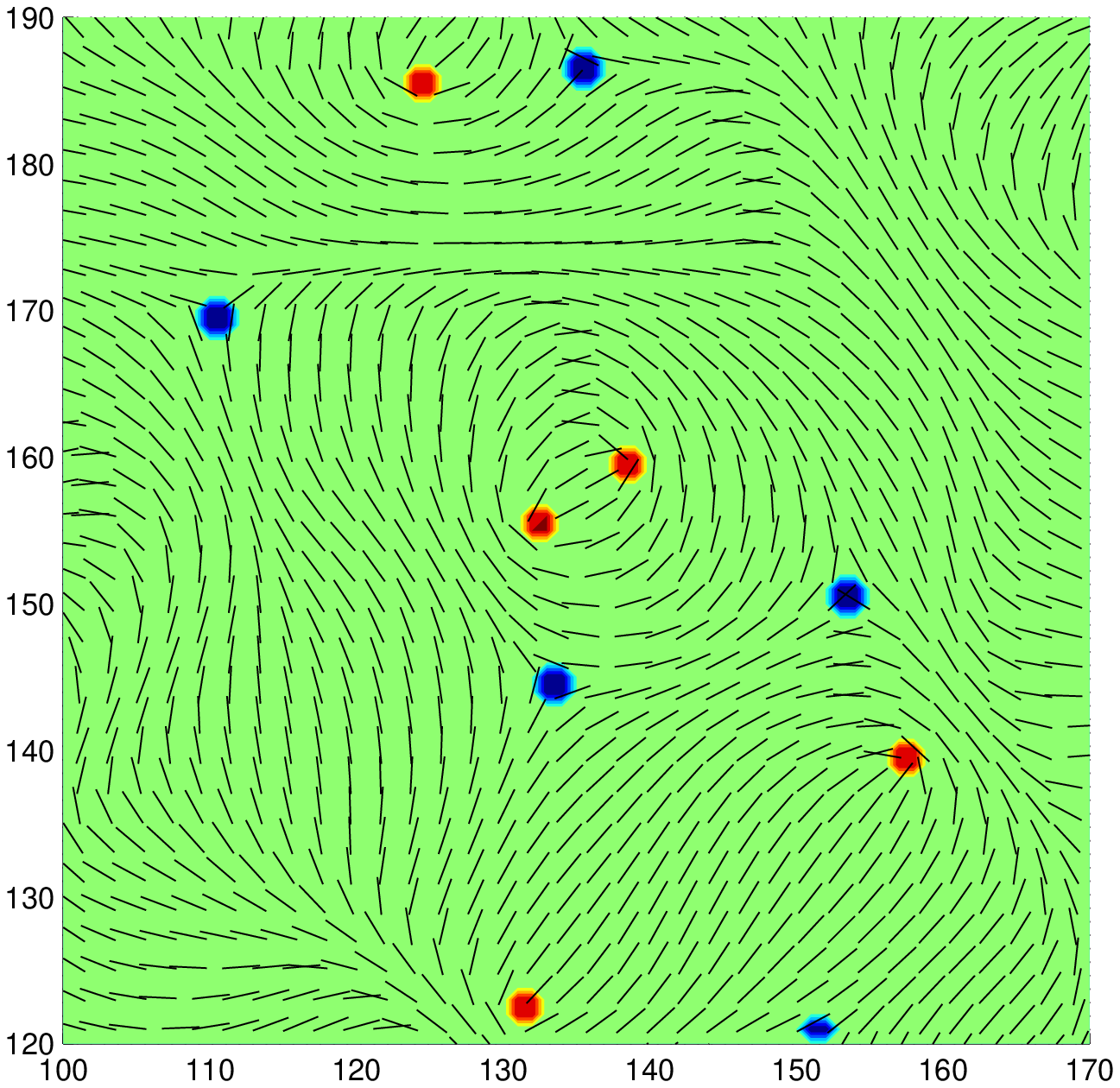}\label{fig:startupcont4}}
\subfigure{\subfigimgloc[trim = 0 0 0 0, clip, width=0.48\linewidth]{\wb{(d)}}{3}{0.25}{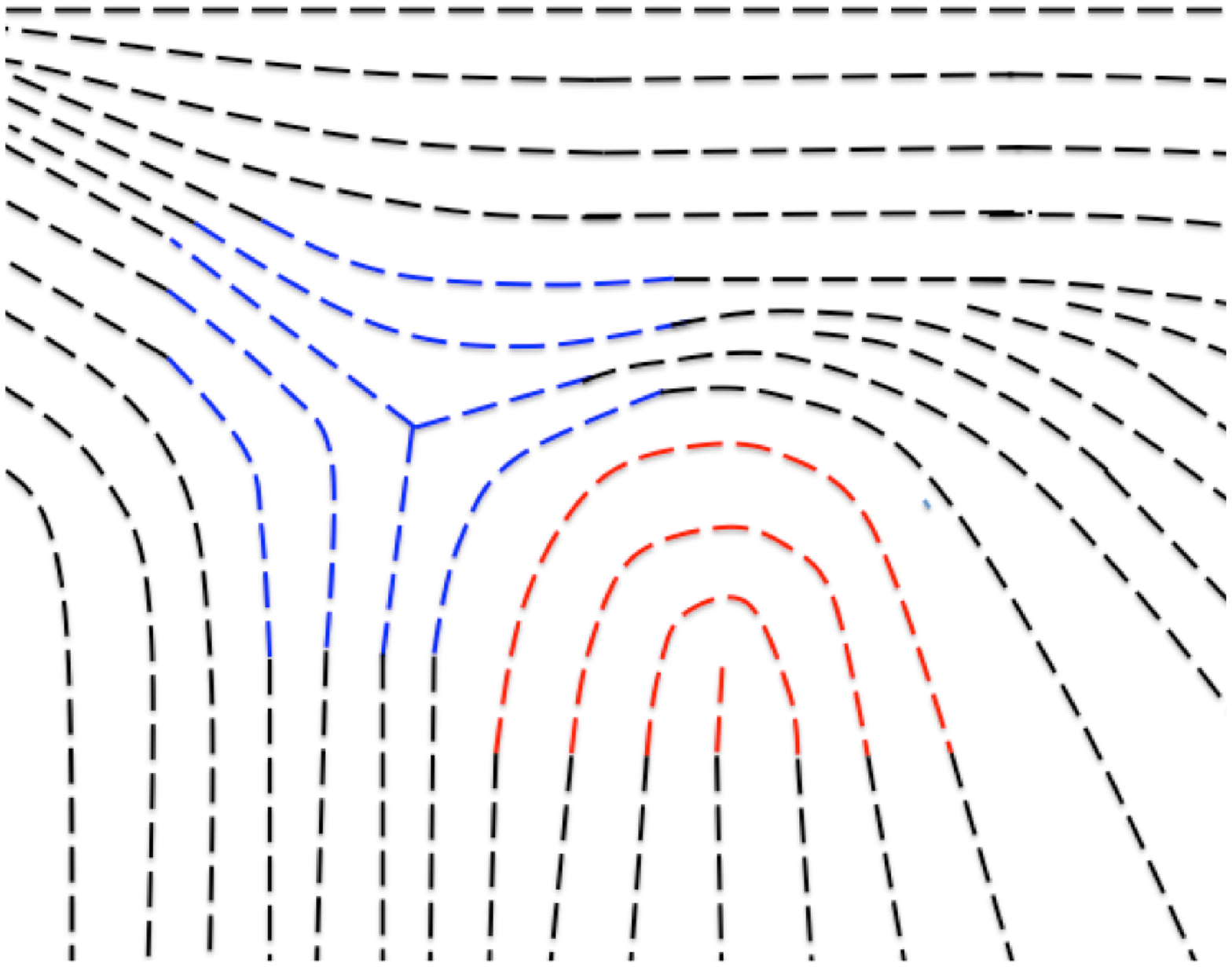}\label{fig:condefects}}
\caption{Snapshots at successive times of the director field and $\pm 1/2$ defects during the development of active turbulence from an ordered nematic state for a contractile system. Two stages are seen: (a),(b) domains of different director orientation are formed, (b),(c) nematic order is restored by the formation of pairs of oppositely charged defects. (d) Local alignment of the director field immediately after the creation of a defect pair.}
\label{fig:startupcon}
\end{figure}
{\em Contractile active nematics}: 
We now comment briefly on contractile systems $\zeta<0$. Here the dominant hydrodynamic instability is to splay deformations. As a result, details of both the initial patterning and of the formation/annihilation of topological defects differ to those observed in the extensile case.   Starting from an orderd configuration as in fig.~\ref{fig:startupt1}, fig.~\ref{fig:startupcon} illustrates the instabilty of  a contractile active nematic. Instead of the bands formed in the extensile case, two dimensional nematic regions of varying orientations appear (fig: \ref{fig:startupcont2}-\ref{fig:startupcont3}). The borders of these regions are marked by large splay deformations. The splay is connected to neighbouring nematic regions through a bend distortion. The bends become more pronounced with time resulting in the formation of pairs of  defects in the border regions as shown in fig.~\ref{fig:startupcont3}-\ref{fig:startupcont4}. The director field immediately after the creation of a defect pair in a contractile nematic  is shown in fig.~\ref{fig:condefects}; compare fig.~\ref{fig:defcre} for the extensile case. 

{\em Summary:} To conclude, the nematic regions in an active system are hydrodynamically unstable. This results in the formation of walls, local lines of high distortion. The elastic energy stored in the walls is released with the creation and annihilation of pairs of defects. Flow both helps to localise the walls and to aid the formation of defects. Defects preferentially move along the walls, but can escape from them at higher activities, and when oppositely charged defects meet they annihilate. Both creation and annihilation events remove walls and help to  reinstate regions of nematic order  which then undergo further hydrodynamic instabilities. The time scale for instability is usually  much faster than the typical time scale of defect dynamics suggesting that it is the creation, motion and annihilation of the defects that primarily control the structure of the director and flow fields. Details of the defect formation differs in contractile suspensions, and further investigations are required to understand the implication of this to the properties of the fully-developed active turbulence.

{\em Acknowledgements:} We thank Z. Dogic, D. Chen and D. Pushkin for helpful discussions. This work was supported by the E.R.C. Advanced Grant MiCE.

\bibliography{refe}

\end{document}